\documentclass[fleqn,usenatbib,useAMS]{mnras}
\usepackage{graphicx}   
\usepackage{amsmath}    
\usepackage{amssymb}    
\usepackage{multicol}        
\usepackage{bm}         
\usepackage{pdflscape}  
\usepackage[T1]{fontenc}
\usepackage{ae,aecompl}
\usepackage{newtxtext,newtxmath}

\def\kms{km s$^{-1}$}

\def\logg{\log \textsl{\textrm{g}}}

\title[Double WDs]{Two new Double-lined Spectroscopic Binary White Dwarfs}
\author[Kilic et al.]
{Mukremin Kilic$^1$,
A. B{\'e}dard$^2$,
P. Bergeron$^2$,
Alekzander Kosakowski$^1$
\\
$^1$Homer L. Dodge Department of Physics and Astronomy, University of Oklahoma, 440 W. Brooks St., Norman, OK, 73019, USA\\
$^2$D\'epartement de Physique, Universit\'e de Montr\'eal, C.P. 6128, Succ. Centre-Ville, Montr\'eal, QC H3C 3J7, Canada\\
}

\date{\ \ Submitted \today \vspace{-0.5cm}}
\pubyear{2019}

\begin{document}
\label{firstpage}
\pagerange{\pageref{firstpage}--\pageref{lastpage}}
\maketitle

\begin{abstract}

We present radial velocity observations of four binary white dwarf candidates identified through their over-luminosity. We identify
two new double-lined spectroscopic binary systems, WD 0311$-$649 and WD 1606+422, and constrain their orbital parameters.
WD 0311$-$649 is a 17.7 h period system with a mass ratio of $1.44 \pm 0.06$ and WD 1606+422 is a  20.1 h period system
with a mass ratio of $1.33 \pm 0.03$. An additional object, WD 1447$-$190, is a 43 h period single-lined
white dwarf binary, whereas WD 1418$-$088 does not show any significant velocity variations over timescales ranging
from minutes to decades. We present an overview of the 14 over-luminous white dwarfs that were identified by B{\'e}dard et al.,
and find the fraction of double- and single-lined systems to be both 31\%. However, an additional
31\% of these over-luminous white dwarfs do not show any significant radial velocity variations. We demonstrate that these must
be in long-period binaries that may be resolved by Gaia astrometry. We also discuss the over-abundance of single low-mass white
dwarfs identified in the SPY survey, and suggest that some of those systems are also likely long period binary systems
of more massive white dwarfs.

\end{abstract}


\section{Introduction}

Double white dwarfs with well-measured orbital and physical parameters provide essential constraints on binary population
synthesis models and the stability of mass transfer in their progenitor systems. However, the number of double
white dwarfs with well measured primary and secondary masses and orbital parameters is rather small, with only 8 eclipsing
and 17 double-lined (SB2) spectroscopic binaries currently known. 

The eclipsing systems are dominated by extremely low-mass (ELM, $M\sim0.2 M_{\odot}$) white dwarfs
\citep[e.g.,][]{steinfadt10,brown11,brown17,burdge19}. This is not surprising,
as the eclipse searches are most sensitive to short period systems, and the formation of ELM white dwarfs requires close binary
companions \citep{li19}. ELM white dwarfs tend to have relatively massive companions \citep{andrews14,boffin15,brown16}, and six of the
known eclipsing systems have relatively large mass ratios. The two exceptions are the low-mass white dwarfs CSS 41177
\citep[0.38 + 0.32 $M_{\odot}$,][]{bours14} and J1152+0248 \citep[0.47 + 0.44 $M_{\odot}$,][]{hallakoun16}.

Double-lined spectroscopic binaries, on the other hand, are biased towards equal brightness and equal mass systems.
The first SB2 white dwarf system, L870$-$2, was identified as a potential binary based on its over-luminosity
compared to other white dwarfs, and was confirmed to be a double-lined spectroscopic binary by \citet{saffer88}.
Radial velocity surveys in the late 1990s and early 2000s, mainly the ESO supernovae type Ia progenitor survey
\citep[SPY,][]{napiwotzki01}, increased the number of SB2 systems to 13
\citep{marsh95,moran97,maxted02,napiwotzki02,karl03,napiwotzki07}. With the serendipitious discovery of an additional system
by \citet{debes15}, and three more by \citet{rebassa17},
the total number of double-lined WDs with well-measured orbital parameters and mass ratios is now 17. There are nine more
SB2 systems identified by \citet{napiwotzki19} that need follow-up radial velocity observations for orbital constraints.

\begin{figure*}
\includegraphics[width=3.0in, bb=0 0 504 360]{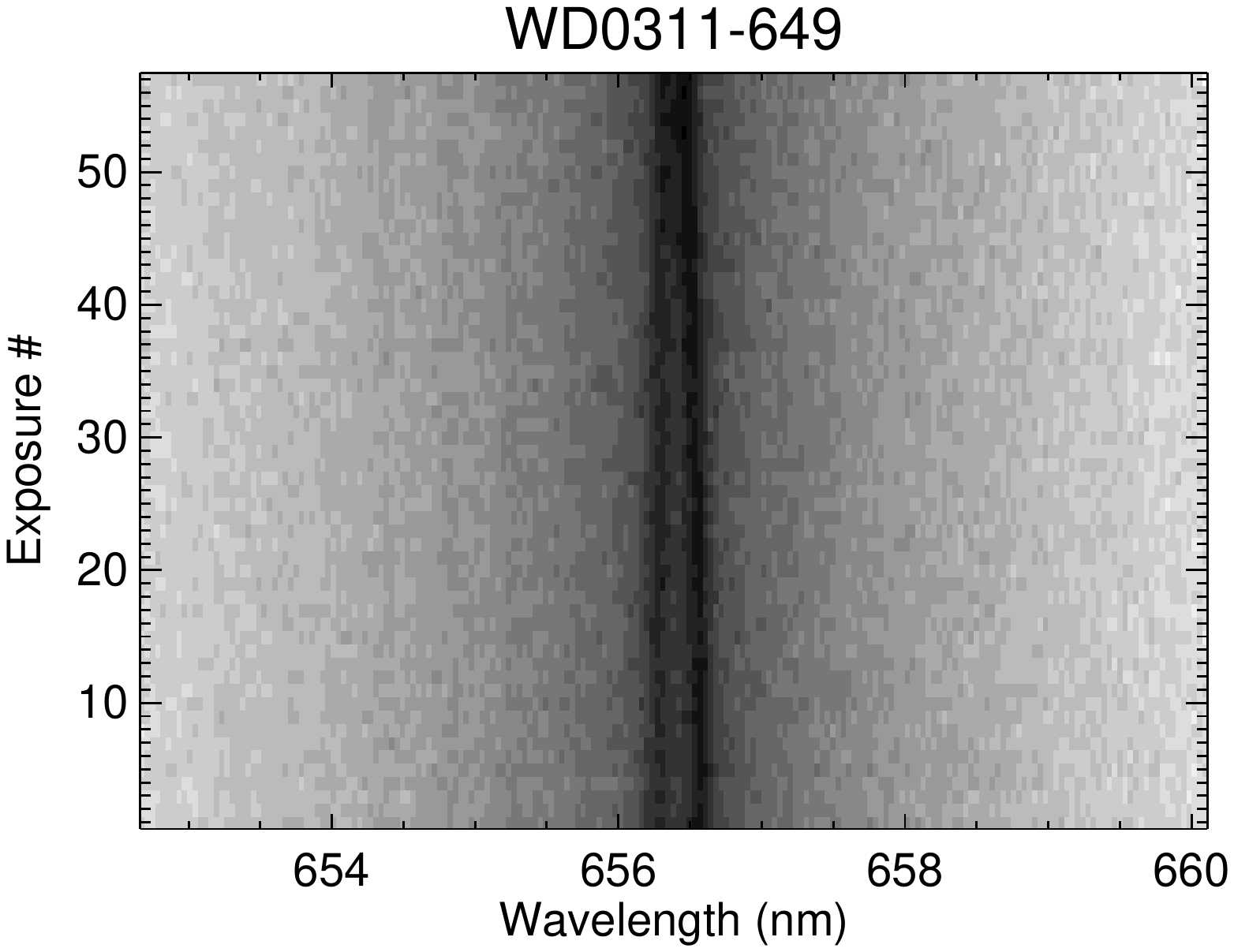}
\includegraphics[width=3.0in, bb=0 0 504 360]{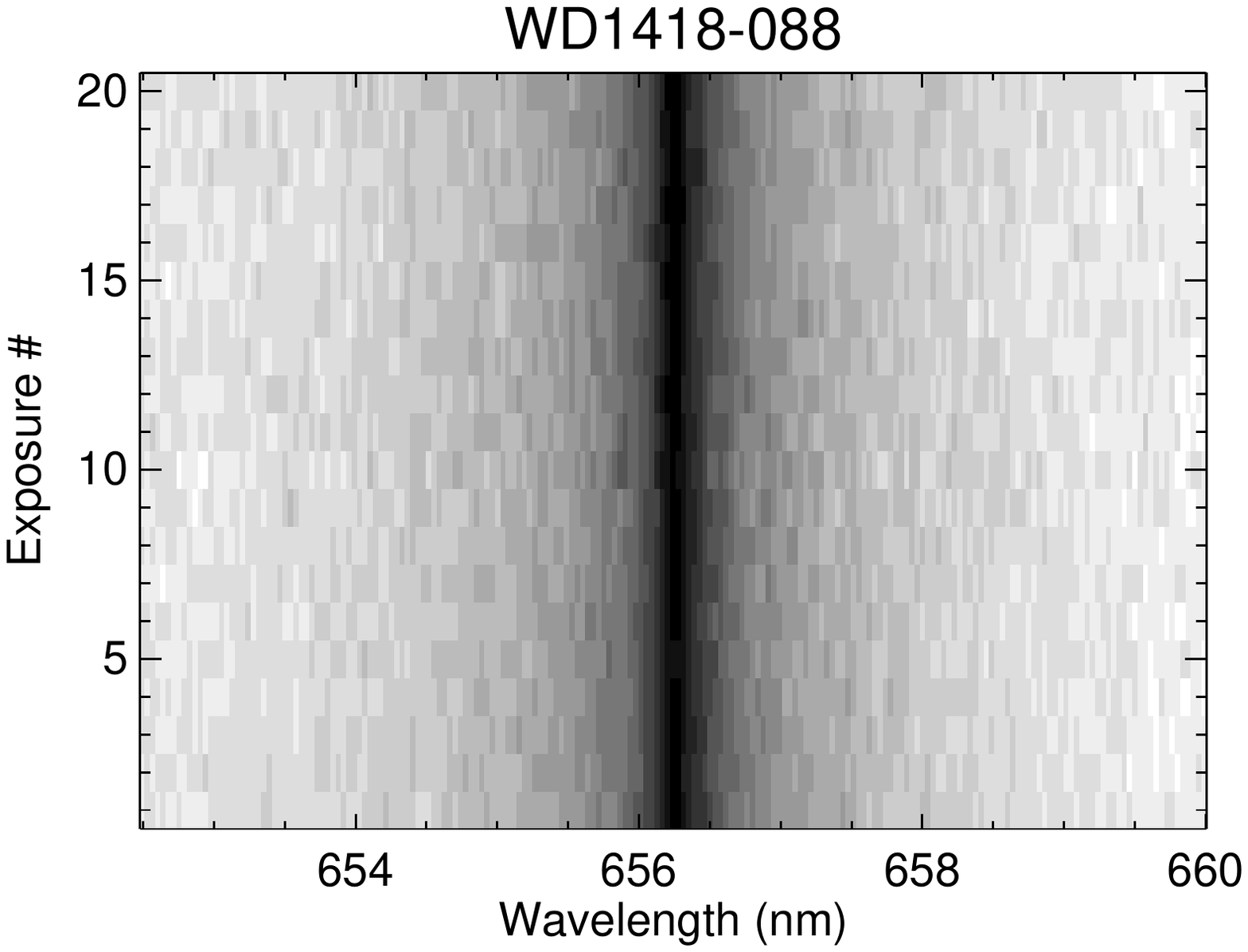}
\includegraphics[width=3.0in, bb=0 0 504 360]{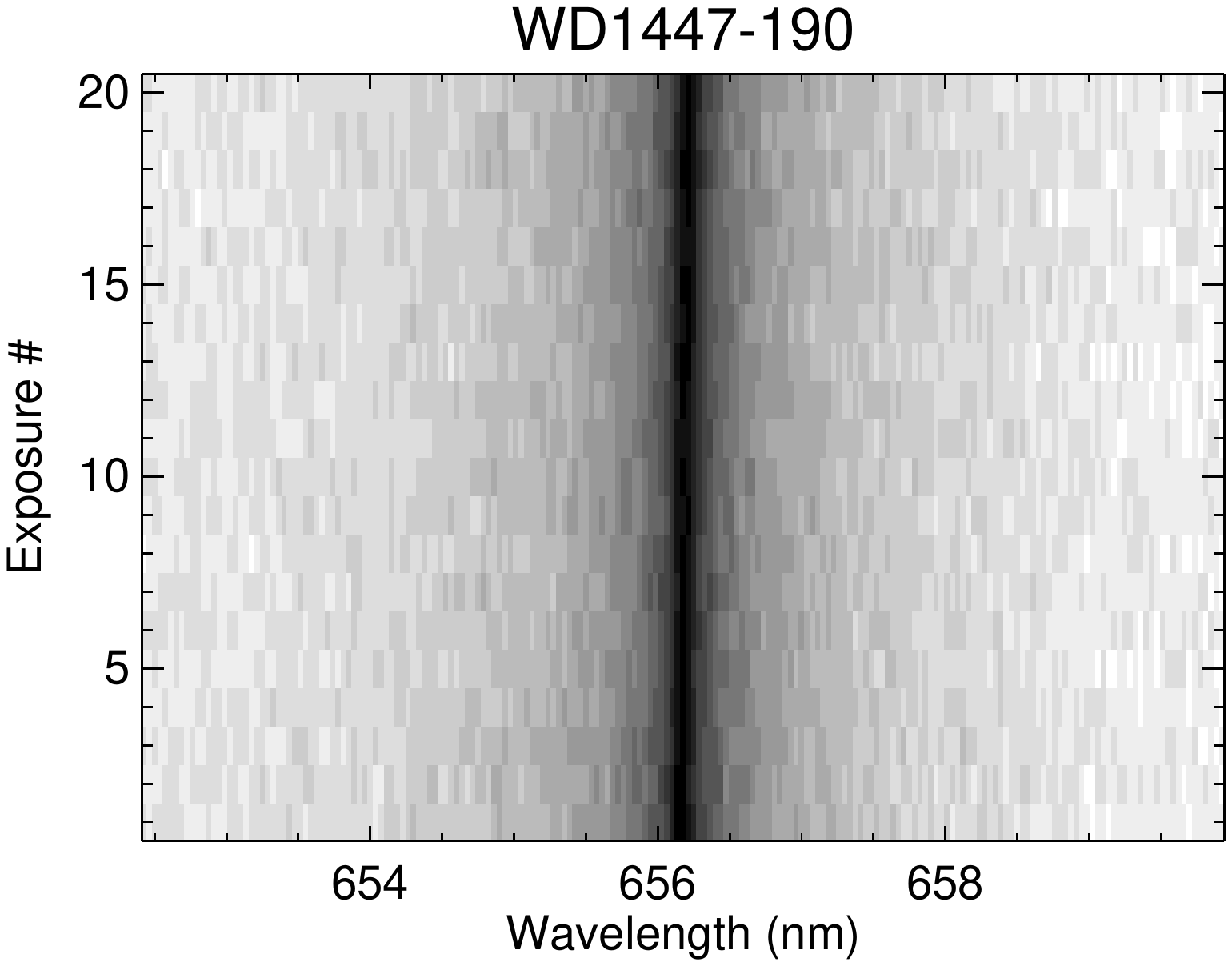}
\includegraphics[width=3.0in, bb=0 0 504 360]{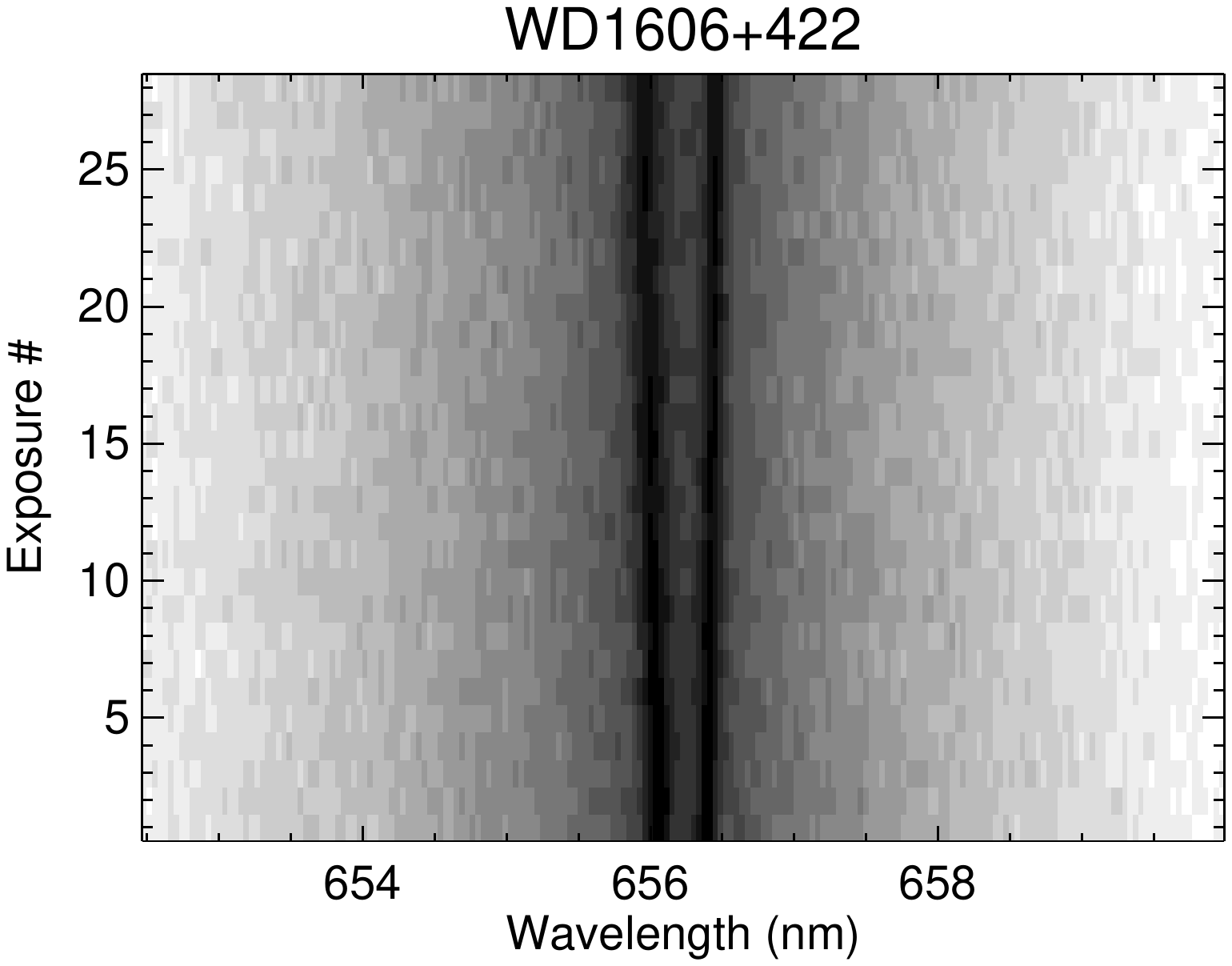}
\caption{Gemini time-resolved spectroscopy of four white dwarfs with back-to-back exposures.
WD 0311$-$649 and WD 1606+422 turn out to be double-lined spectroscopic
binaries, whereas WD 1447$-$190 is a single-lined binary. The remaining target, WD 1418$-$088, does not show any significant
radial velocity variability.}
\label{fig1}
\end{figure*}

Recently, \citet{bedard17} performed a spectroscopic and photometric analysis of 219 white dwarfs with trigonometric parallax
measurements available at the time, and identified 15 unresolved double degenerate binary candidates, including several previously known
double-lined spectroscopic binaries. \citet{bergeron19} presented an updated analysis of these objects based on Gaia Data Release
2 parallaxes, and confirmed the over-luminous nature of all but one of these targets, WD1130+189. 
Here we present follow-up spectroscopy of four of these white dwarfs, three of
which are confirmed to be binary systems. We present an overview of the over-luminous white dwarf sample from \citet{bedard17}
and discuss the period and mass distribution of the sample, as well as the fractions of SB2, SB1, and systems that show no significant
radial velocity variability.

\section{Observations}

We used the HIRES echelle spectrometer \citep{vogt94} on the Keck I telescope at Mauna Kea Observatory in Hawaii to
observe three of our targets on UT 2018 June 18. Unfortunately our three half-night long observing run was limited to a
period of about only 2 hours due to volcanic activity and vog. We used the blue cross disperser with a 1.15 arcsec slit
resulting in a spectral resolution of 37,000. We used {\sc MAKEE} to analyze the HIRES data.

We obtained follow-up optical spectroscopy of all four of our targets using the 8m Gemini telescopes equipped 
with the Gemini Multi-Object Spectrograph (GMOS) as part of the queue programs GN-2018A-Q-116 and
GS-2018B-Q-117. We used the R831 grating and a
0.25$\arcsec$ slit, providing wavelength coverage from 5380 \AA\ to 7740 \AA\ and a resolution of 0.376 \AA\ per pixel.
Each spectrum has a comparison lamp exposure taken within 10 min of the observation time. We used the {\sc IRAF GMOS}
package to reduce these data. 

Our initial observing strategy included a series of back-to-back exposures to look for short period systems. 
Figure \ref{fig1} shows the Gemini/GMOS trailed spectra for all four targets based on these back-to-back exposures. We obtained
$57\times150$ s exposures of WD 0311$-$649 on UT 2018 Oct 2, $20\times300$ s exposures of WD 1418$-$088 on
UT 2018 July 1, $20\times300$ s exposures of WD 1447$-$190 on UT 2018 July 10, and $28\times245$ s exposures of
WD 1606+422 on UT 2018 Sep 11. This figure reveals two double-lined binary systems, WD 0311$-$649
and WD 1606+422, where the double H$\alpha$ lines are seen converging and diverging over a few hours, respectively.
One of the single-lined objects, WD 1447$-$190, also showed significant velocity shifts in the back-to-back exposures,
but the other, WD 1418$-$088, did not show any significant variations over a period of 1.8 h, and we decided not to follow
it up further (see more below). To constrain the orbital parameters of the three velocity variable systems, we obtained additional
spectroscopy with different nightly cadences as part of the Gemini Fast Turnaround and queue programs
GN-2019A-FT-208, GS-2019A-FT-202, and GS-2019B-Q-113.

We obtained seven additional spectra of WD 1447$-$190 on UT 2019 March 1-3 at the 4.1m SOAR telescope equipped with
the Goodman High Throughput spectrograph \citep{clemens04} with the 930 line mm$^{-1}$ grating and the 1.03$\arcsec$ slit.
This set-up provides 2.2 \AA\ spectral resolution over the range $3550-5250$ \AA. The SOAR spectra were obtained as part of the
NOAO program 2019A-0134.

\section{Radial Velocity Measurements}

\begin{figure}
\centering
\includegraphics[width=3in]{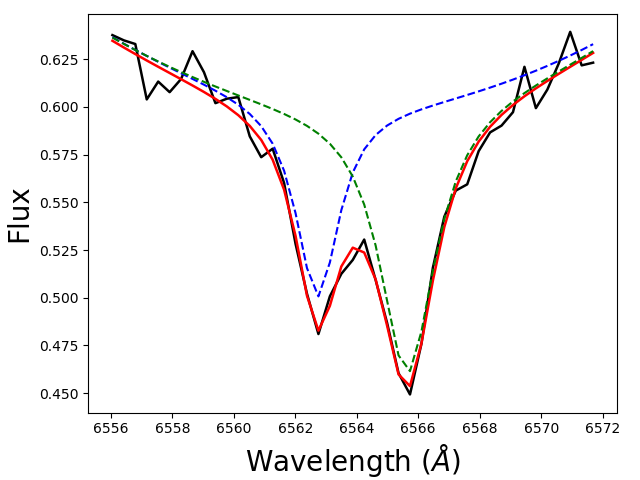}
\caption{Best-fitting Lorentzian profiles to the H$\alpha$ line cores visible in the double-lined spectroscopic binary
WD 0311$-$649 (blue and green dotted lines). The red solid line shows the composite best-fitting model.}
\label{fig2}
\end{figure}

We use the core of the H$\alpha$ line to measure the radial velocities of our single- and double-lined systems.
After normalizing the continuum, we use a quadratic polynomial plus a Lorentzian or Voigt profile to fit the line wings and
the line cores, respectively. We find the best-fit parameters with LMFIT, a version of the Levenberg-Marquardt algorithm adapted for
Python \citep{newville14}. We apply the standard Solar System barycentric corrections, and use the night skylines to check the
spectrograph flexure. 

Figure \ref{fig2} shows the best-fit to the first Gemini exposure on the double-lined system WD 0311$-$649, demonstrating
our procedure. Here the dotted blue and green lines show the best-fitting Lorentzian profiles to the two H$\alpha$ line cores,
and the red line shows the composite best-fitting model. The formal measurement errors on the two H$\alpha$ line centers
in this spectrum are 0.06 and 0.05 \AA, respectively. \citet{napiwotzki19} demonstrated that formal fitting errors tend to
be underestimated, and that error estimates based on bootstrapping are better for including uncertainties from imperfections
of the input data and non-Gaussian noise. We use a similar bootstrapping procedure, and randomly select $N$ points of
the observed spectra, where points can be selected more than once. We use the bootstrapped spectra
to rederive velocities, repeating this procedure 1000 times. We add the standard deviation of the velocity measurements
from the bootstrapped spectra and the formal fitting errors in quadrature to estimate the total errors in each velocity
measurement. This procedure gives 4 \kms\ errors for both lines in the spectrum shown in Figure \ref{fig2}.

\begin{figure}
\centering
\includegraphics[width=3in, bb=54 54 758 558]{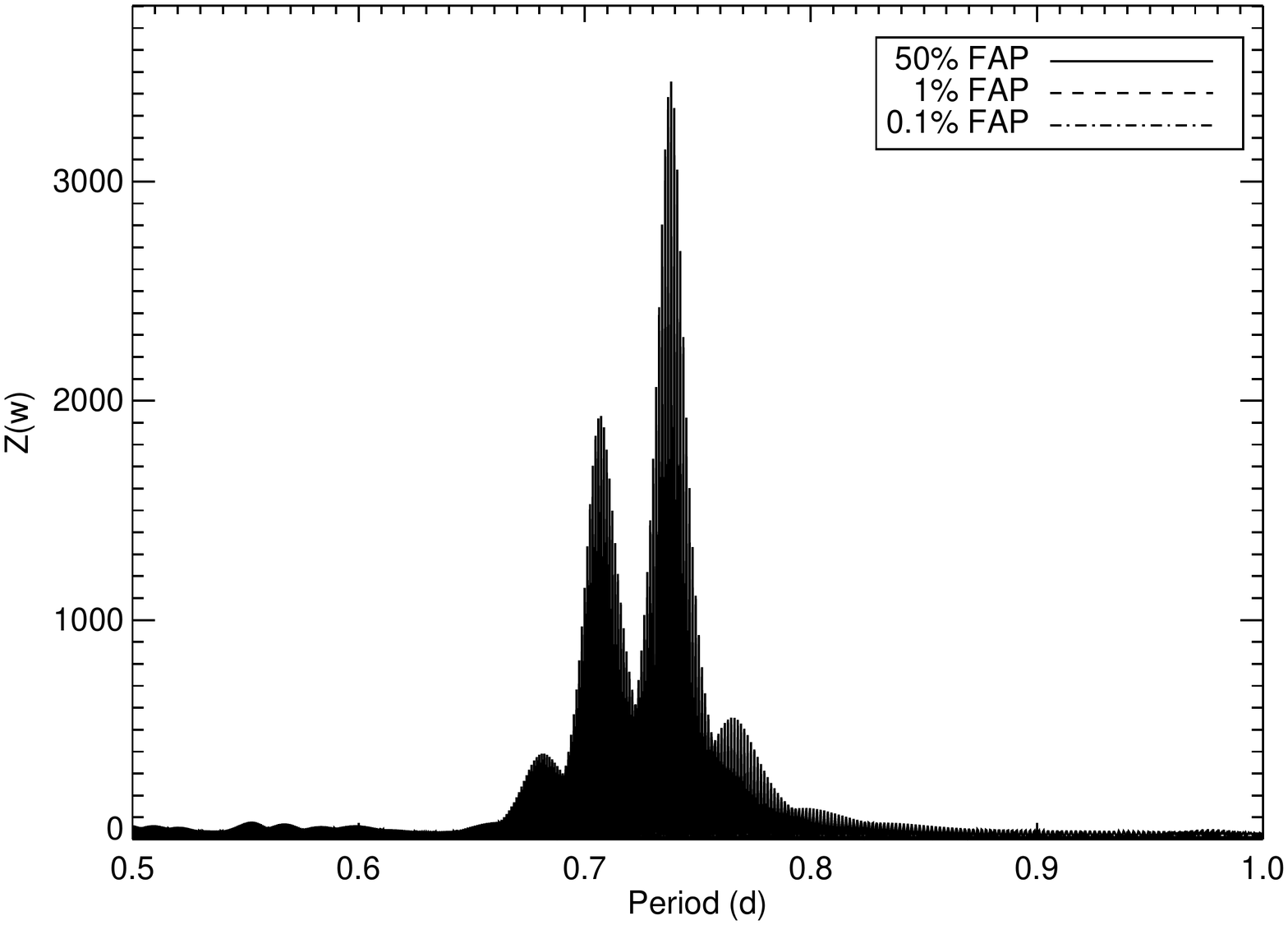}
\includegraphics[width=3in, bb=18 144 592 718]{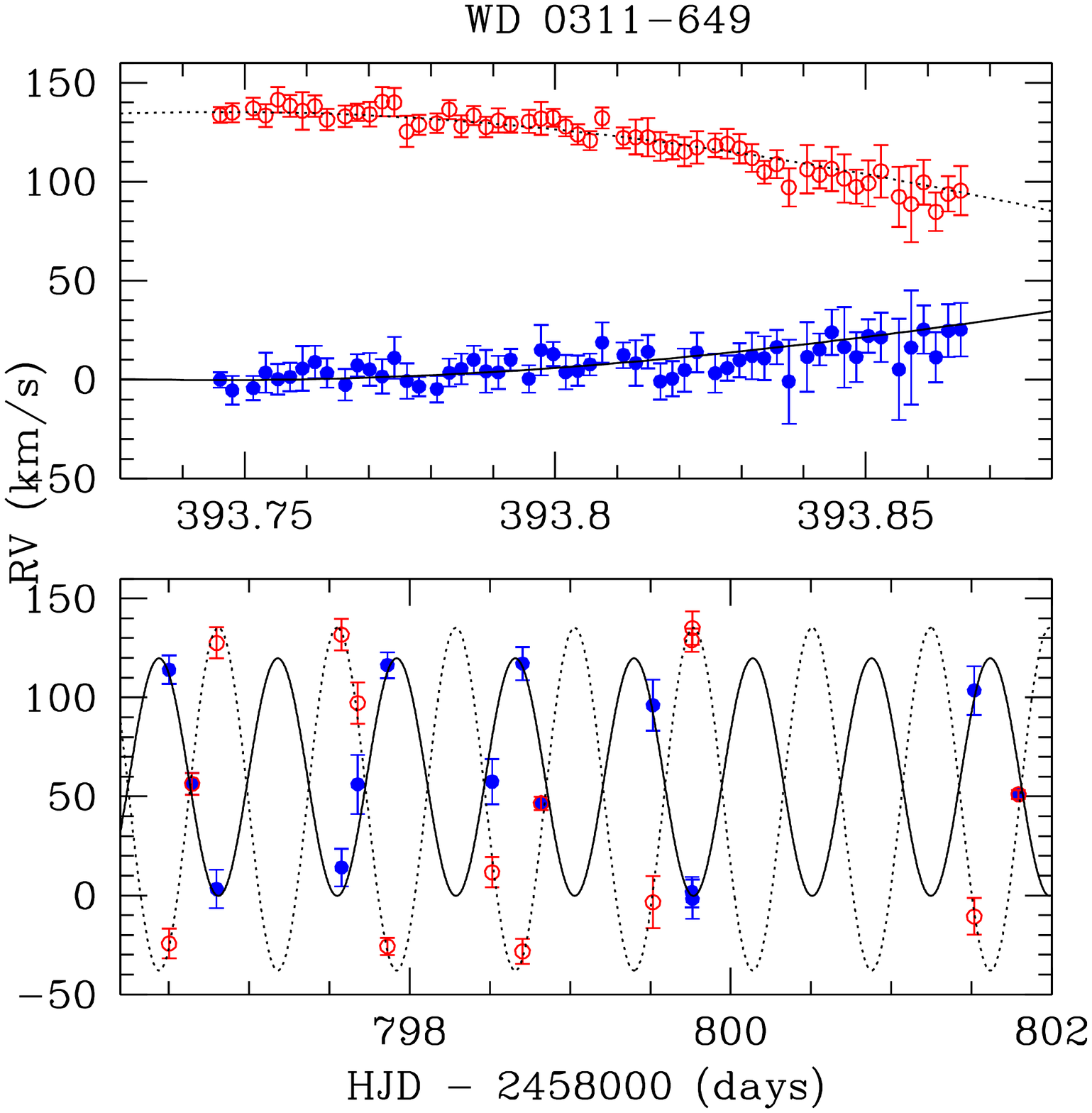}
\caption{{\it Top:} Lomb-Scargle periodogram for WD 0311$-$649. 
{\it Middle and Bottom:} Radial velocity measurements (open and filled points) and
the best-fitting orbital solutions (dotted and solid lines) for the two stars in WD 0311$-$649
assuming a circular orbit.}
\label{fig:0311}
\vspace{-0.2in}
\end{figure}

\section{Atmospheric Parameter Determination}

\citet{bedard17} showed that it is possible to constrain the atmospheric parameters ($T_{\rm eff}$ and $\logg$) of both white dwarfs
in an unresolved DA+DA binary system by combining spectroscopic, photometric, and astrometric information. More specifically, they
developed a deconvolution procedure that involves fitting simultaneously the observed Balmer lines and spectral energy distribution
with composite model atmospheres. We briefly describe this method here, as we apply it in Section \ref{sec:results} to revisit the
atmospheric properties of our four binary candidates in light of our new data.

The radiative flux $f_{\nu}$ received at Earth from an unresolved double degenerate system is simply the sum of the Eddington fluxes
emitted by the individual components, properly weighted by their respective solid angle:
\begin{equation}
f_{\nu} = 4 \pi \left( \frac{R_1}{D} \right)^2 H_{\nu,1} + 4 \pi \left( \frac{R_2}{D} \right)^2 H_{\nu,2} \ .
\label{eq:flux}
\end{equation}
Assuming that the distance $D$ is known from a trigonometric parallax measurement, the right-hand side of this equation depends
only on the four atmospheric parameters $T_{\rm eff,1}$, $\logg_1$, $T_{\rm eff,2}$, and $\logg_2$. Indeed, for given values of these
quantities, the Eddington fluxes $H_{\nu,1}$ and $H_{\nu,2}$ are obtained from model atmospheres, and the radii $R_1$ and $R_2$ are
obtained from evolutionary sequences. In what follows, we use pure-hydrogen model atmospheres similar to those described in
\citet{tremblay09} together with 3D hydrodynamical corrections from \citet{tremblay13}, as well as evolutionary models similar to
those described in \citet{fontaine01} with carbon/oxygen cores ($X_C = X_O = 0.5$) and standard ``thick'' hydrogen layers
($M_{\rm H}/M_{\star}=10^{-4}$).

Both our spectroscopic and photometric analyses are based on Equation \ref{eq:flux}. In the spectroscopic case, the observed optical
spectrum (left-hand side) is compared to a weighted sum of two synthetic spectra (right-hand side). Since only the shape of the
Balmer lines is of interest, the observed and (combined) synthetic spectra are normalized to a continuum set to unity before the
comparison is carried out. In the photometric case, a set of average fluxes measured in some optical and infrared
bandpasses (left-hand side) is compared to a weighted sum of two synthetic spectra properly averaged over the corresponding
bandpass filters (right-hand side). We adopt the zero points given in \citet{holberg06} to convert observed magnitudes into average
fluxes, and neglect reddening since all of our systems are within 50 pc (see Table 1). Note that absolute fluxes are required here since we are interested in the overall energy distribution. 

Our fitting procedure uses the Levenberg-Marquardt algorithm to find the values
of $T_{\rm eff,1}$, $\logg_1$, $T_{\rm eff,2}$, and $\logg_2$ that minimize the difference between the two sides of
the equation for both the spectroscopic and photometric observations simultaneously. We stress that such a unified
approach is mandatory to achieve reliable results, especially when all four atmospheric parameters are allowed to vary.
 We do not fit for distance, since distances are precisely determined by Gaia DR2 \citep{gaia18} for all four targets (Table 1).

In our analysis reported below, we employ the same spectroscopic and photometric data as in \citet{bedard17}, with the exceptions
that we make use of our new SOAR spectrum for WD 1447$-$190, and that we add the Pan-STARRS $grizy$ magnitudes \citep{chambers16}
to our photometric fits for WD 1418$-$088, WD 1447$-$190, and WD 1606+422. 

\begin{figure*}
\includegraphics[width=2.9in,angle=270,clip=true,trim=2.2in 0.0in 1.6in 0.0in]{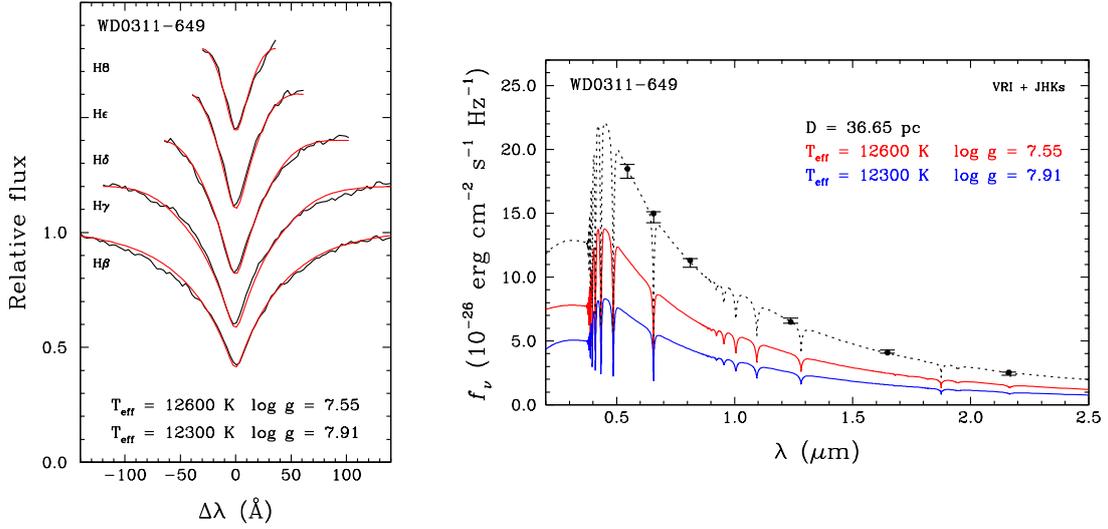}
\caption{Best model-atmosphere fit to the Balmer lines (left panel) and the spectral energy distribution (right panel) of WD 0311$-$649. In 
the left panel, the observed and synthetic spectra are displayed as the black and red lines, respectively. In the right panel,
the observed and synthetic average fluxes are shown as the error bars and filled circles, respectively; in addition, the red and
blue lines show the contribution of each component to the total monochromatic model flux, which is displayed as the black dotted
line. The best-fitting atmospheric parameters are given in both panels.}
\label{fig:fit0311}
\end{figure*}

\begin{figure}
\includegraphics[width=3.4in,clip=true,trim=1.7in 4.2in 1.7in 3.8in]{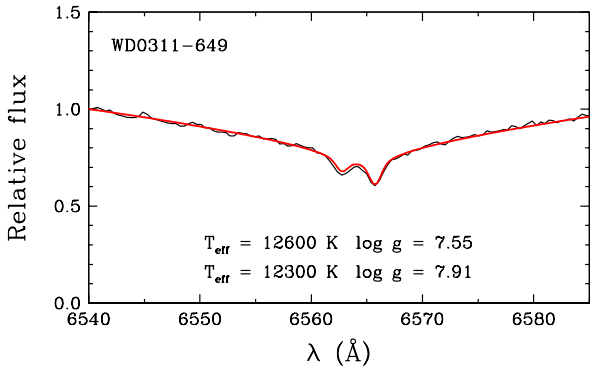}
\caption{Comparison of the observed double H$\alpha$ feature of WD 0311$-$649, shown as the black line, with that predicted by 
our best model-atmosphere fit, displayed as the red line.}
\label{fig:Ha0311}
\end{figure}

\section{Results}
\label{sec:results}
\subsection{WD 0311$-$649}

Figure \ref{fig:0311} shows the radial velocity measurements and the Lomb-Scargle periodogram for
the double-lined spectroscopic binary WD 0311$-$649. We use the IDL program MPRVFIT \citep{delee13}
in the SB2 mode to find the best-fitting orbit. Excluding the three spectra
where the H$\alpha$ lines from both stars overlap and appear as a single line, we have 68 radial velocity
measurements for each star. The best-fitting orbital parameters and their formal errors are
$P = 0.73957 \pm 0.00001$ d, $K_1 = 86.5 \pm  1.7$  \kms, $K_2 = 60.0 \pm  2.1$ \kms,
$\gamma_1 = 48.7 \pm 1.7$ \kms,
and a velocity offset of $\gamma_2 - \gamma_1 = 11.1 \pm 2.7$ \kms. 

The Lomb-Scargle diagram shows that there are significant aliases, which are offset from each other by multiples of
0.00135 d. Period aliases are the largest source of uncertainty in our orbital solutions.
To constrain the impact of these aliases on our orbital solutions, we use a Monte-Carlo approach, re-sampling
the radial velocities with their errors and re-fitting orbital parameters 1,000 times. This approach samples
$\chi^2$ space in a self-consistent way. We report the median value and errors derived from the 15.9\% and 84.1\%
percentiles of the distributions for each orbital element. The best-fitting orbital parameters from the Monte Carlo
simulations are $P = 0.73956^{+0.00134}_{-0.00267}$ d, $K_1 = 86.5^{+2.0}_{-1.7}$ \kms,
$K_2 = 60.1^{+2.0}_{-2.1}$ \kms, $\gamma_1 = 48.4 \pm 1.7$ \kms,
$\gamma_2 - \gamma_1 = 11.6^{+2.6}_{-2.7}$ \kms, and $\frac{K_1}{K_2} = 1.44 \pm 0.06$.
These values are consistent with the formal estimates from MPRVFIT within the errors, though the error in period
is significantly larger than the formal errors due to the aliasing present. 

The individual masses of the two components can be derived from the orbital parameters. Since the difference
in systemic velocities is equal to the difference in gravitational redshifts, we have:
\begin{equation}
\begin{split}
\gamma_2 - \gamma_1 & = \frac{G}{c} \left( \frac{M_2}{R_2(M_2)} - \frac{M_1}{R_1(M_1)} \right) \\
& = \frac{G}{c} \left( \frac{K_1M_1/K_2}{R_2(K_1M_1/K_2)} - \frac{M_1}{R_1(M_1)} \right)
\end{split}
\label{eq:mass}
\end{equation}
where $R(M)$ is the mass-radius relation obtained from our evolutionary sequences. For given values of $\gamma_2 - \gamma_1$
and $K_1/K_2$, this equation can be solved numerically for $M_1$ (and hence $M_2$). We find $M_1 = 0.385^{+0.060}_{-0.063} \ M_{\odot}$
and $M_2 = 0.554^{+0.073}_{-0.082} \ M_{\odot}$.

Figure \ref{fig:fit0311} displays our best model-atmosphere fit to the Balmer lines and the spectral energy distribution of
WD 0311$-$649. In the minimization procedure, the surface gravities are held fixed to the values derived from the orbital
solution, $\logg_1 = 7.55^{+0.14}_{-0.17}$ and $\logg_2 = 7.91^{+0.12}_{-0.16}$, so only the effective temperatures are treated as
free parameters. Our fitting method yields $T_{\rm eff,1} = 12,600 \pm 500$ K and $T_{\rm eff,2} = 12,300 \pm 500$ K. Both the
spectroscopic and photometric data are nicely reproduced by our composite model. 

To validate our solution further, we compare in Figure \ref{fig:Ha0311} the
observed and predicted H$\alpha$ features of WD 0311$-$649. Five of our Gemini spectra are co-added in order to increase the
signal-to-noise ratio, and a wavelength shift is applied to the individual synthetic spectra to match the observed shift
between the two line cores. The agreement is almost perfect, even though no fit was performed here, and thus confirms the
accuracy of our atmospheric parameters. Finally, it is interesting that our solution places the secondary star
within the ZZ Ceti instability strip, close to the blue edge \citep{gianninas11}. Therefore, this object should be monitored for
pulsations, which might however be difficult to detect due to the light of the primary star.

\subsection{WD 1418$-$088}

\begin{figure}
\centering
\includegraphics[width=3in, bb=18 144 592 718]{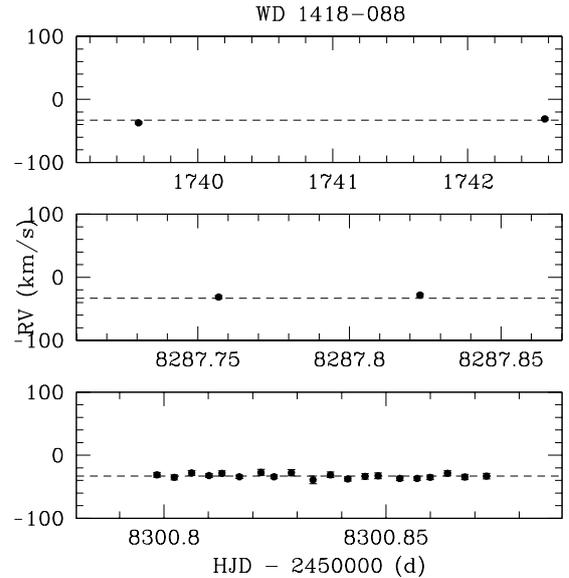}
\caption{VLT (top), Keck (middle), and Gemini (bottom) radial velocity observations of WD 1418$-$088. The dotted line marks
the weighted mean of the velocity measurements.}
\label{fig:1418}
\end{figure}

Figure \ref{fig:1418} shows the radial velocity measurements for WD 1418$-$088 from the SPY survey \citep[top panel,][]{napiwotzki19}
and our Keck (middle panel) and Gemini (bottom panel) observations. WD 1418$-$088 does not show any significant radial velocity
variations. We use the weighted mean velocity to calculate the $\chi^2$ statistic for a constant velocity model. The probability, $p$, of
obtaining the observed value of $\chi^2$ or higher from random fluctuations of a constant velocity, taking into account the appropriate
number of degrees of freedom is 0.22. Hence, the null hypothesis cannot be rejected; 
the radial velocity measurements for WD 1418$-$088 are consistent with a constant velocity.

\citet{bedard17} showed that there is  a significant discrepancy between the spectroscopic and photometric solutions for
WD 1418$-$088. Under the assumption of a single star, the spectroscopic fits (corrected for 3D effects) indicate
$T_{\rm eff} = 8060$ K,  $\log{g} = 8.1$,  and $M = 0.66 \ M_{\odot}$, whereas the photometric fits using the Gaia DR2 parallax
measurement indicate much lower $\log{g}=7.55$ and $M=0.36 \ M_{\odot}$ \citep{blouin19}. The only way to resolve
the discrepancy between the photometric and spectroscopic solutions is if WD 1418$-$088 is a binary system.
The combined light from the two stars in an unresolved binary would make it appear more luminous, which
could be interpreted as a single white dwarf having a larger radius, and therefore a lower mass.

\begin{figure}
\centering
\includegraphics[width=2.8in, bb=18 144 592 718]{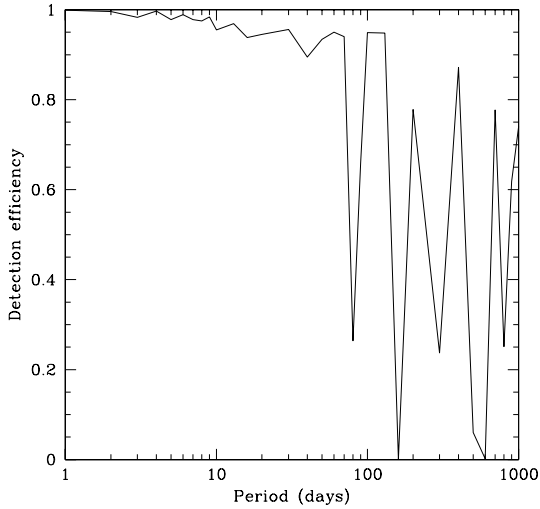}
\caption{The detection efficiency as a function of orbital period for WD 1418$-$088, assuming an equal mass binary system.}
\label{fig:det}
\end{figure}

Given the limited number of radial velocity observations over monthly and yearly timescales, our observations of WD 1418$-$088
are not sensitive to long period systems. To estimate the detection efficiency of a binary system as a function of orbital period,
we use a Monte Carlo approach, and generate synthetic radial velocity measurements with the same temporal sampling and
accuracy as the WD 1418$-$088 observations. We assume a mass ratio of one, and include the projection effects due to
randomly oriented orbits. We estimate our detection efficiency using the number of trials which satisfy the detection criterion of
$\log{(p)} < -4$ \citep[see][]{maxted00}. Figure \ref{fig:det} shows the detection efficiency of our observations for WD 1418$-$088
for orbital periods ranging from 1 to 1000 days. This figure shows that we would have detected the majority of the
binary systems with orbital periods $\leq70$ days, but our detection efficiency significantly deteriorates beyond 80 days.
Hence, WD 1418$-$088 is likely a long period binary white dwarf system. One of the over-luminous white dwarfs included in the
\citet{bedard17} study, WD 1639+153, is an astrometric binary with an orbital period of 4 years. Hence, WD 1418$-$088 may
be an unresolved binary with a similarly long orbital period.

\begin{figure*}
\includegraphics[width=2.9in,angle=270,clip=true,trim=2.2in 0.0in 1.6in 0.0in]{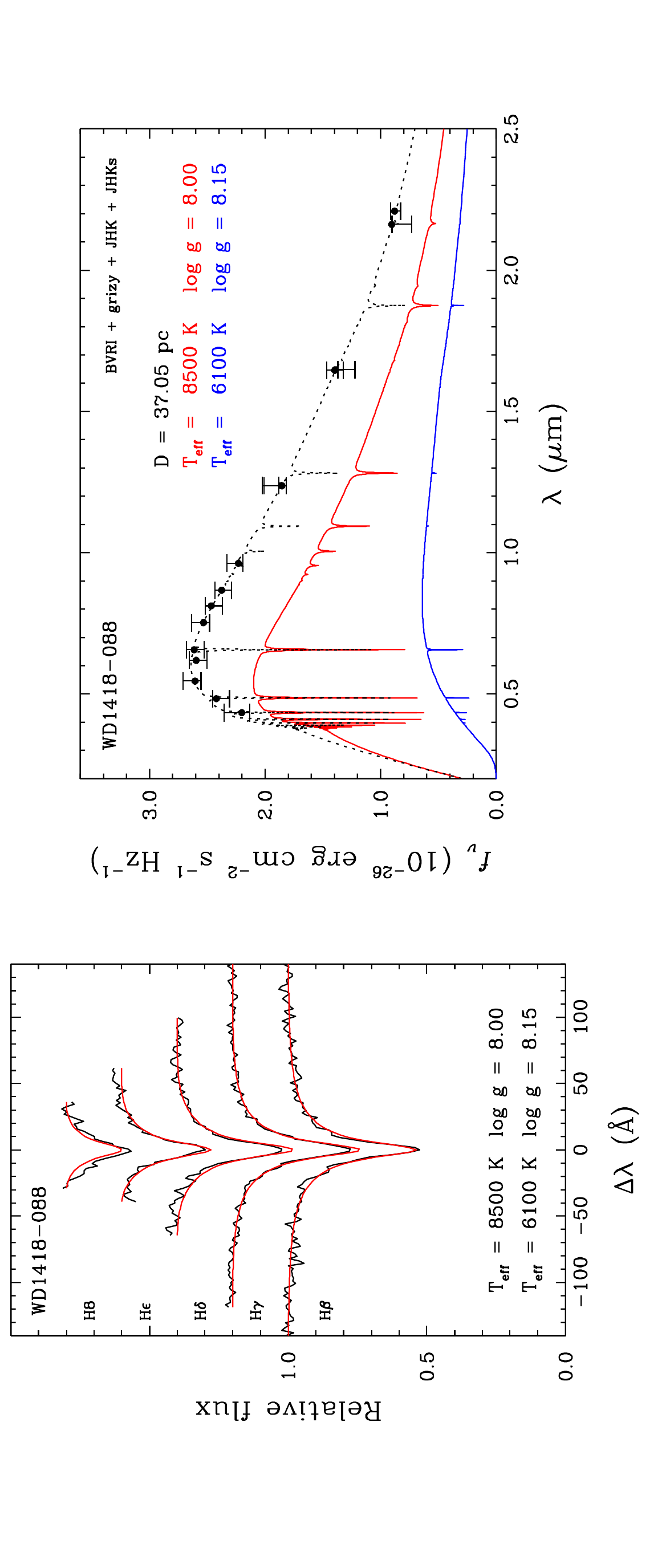}
\includegraphics[width=2.9in,angle=270,clip=true,trim=2.1in 0.0in 1.6in 0.0in]{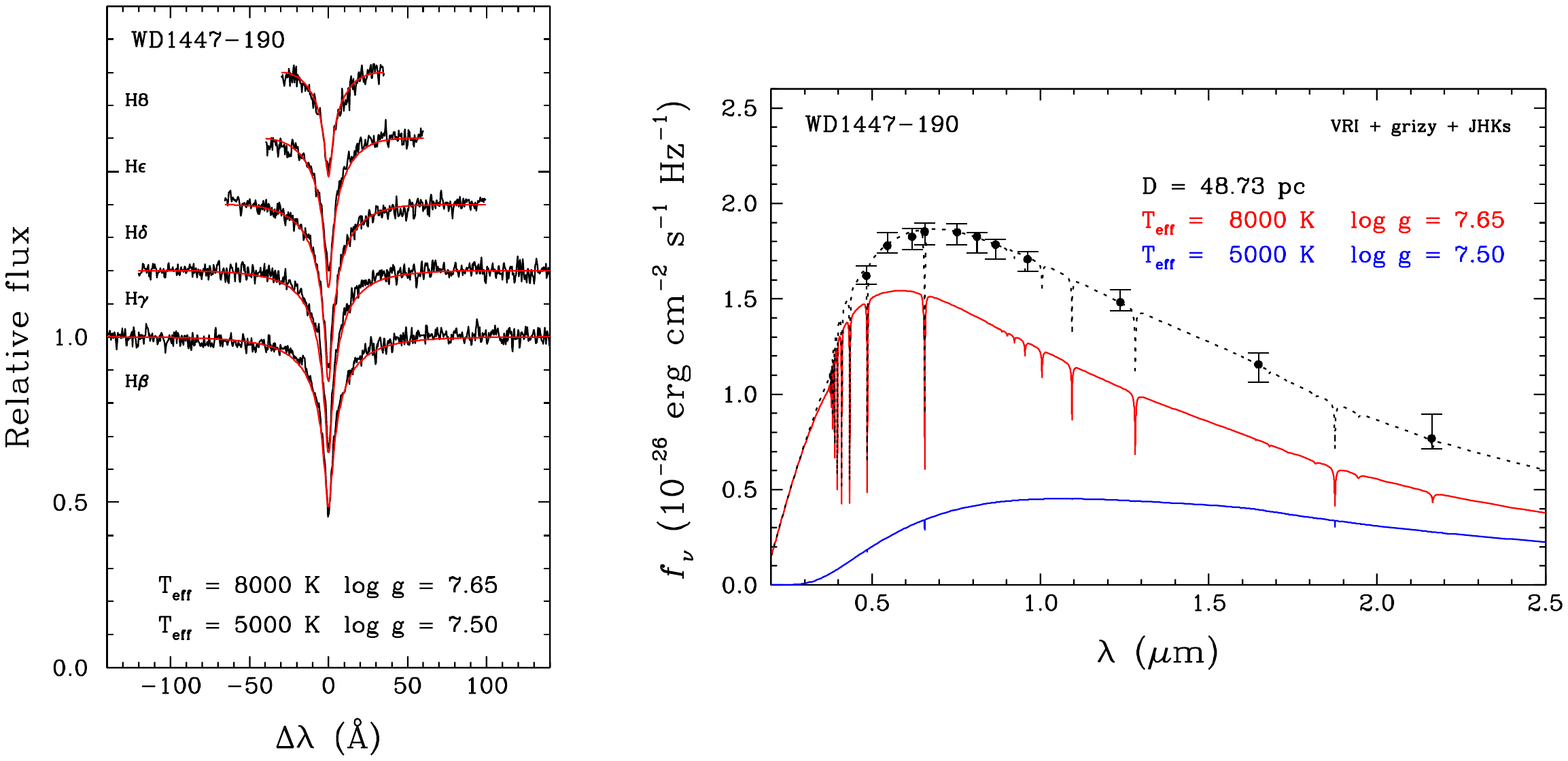}
\includegraphics[width=2.9in,angle=270,clip=true,trim=2.2in 0.0in 1.6in 0.0in]{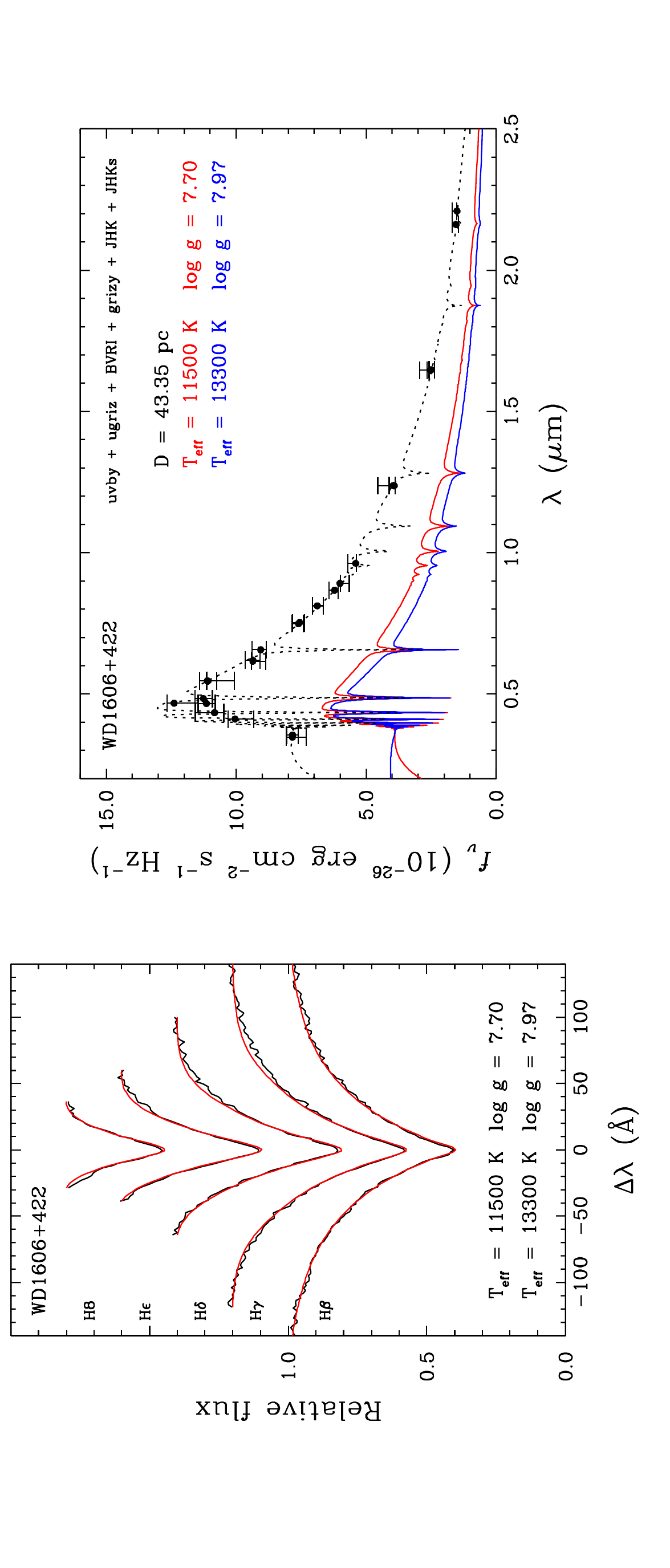}
\caption{Same as Figure \ref{fig:fit0311}, but for WD 1418$-$088, WD 1447$-$190, and WD 1606+422.}
\label{fig:fit3}
\end{figure*}

Figure \ref{fig:fit3} displays our best model-atmosphere fit to the spectroscopic and photometric observations
of WD 1418$-$088 (as well as WD 1447$-$190 and WD 1606+422) assuming a DA+DA binary system. Here,
in the absence of an orbital solution, all four atmospheric parameters are
allowed to vary in the fitting procedure and are thus less tightly constrained than in the case of WD 0311$-$649. We derive
$T_{\rm eff,1} = 8500 \pm 1000$ K, $\logg_1 = 8.00 \pm 0.20$ and $T_{\rm eff,2} = 6100 \pm 500$ K, $\logg_2 = 8.15 \pm 0.20$, which
correspond to $M_1 = 0.60^{+0.12}_{-0.11} \ M_{\odot}$ and $M_2 = 0.68^{+0.13}_{-0.12} \ M_{\odot}$ according to the mass-radius relation.
The agreement between the data and the model is excellent, but the errors in our surface gravity and mass estimates are relatively
large due to the lack of any orbital constraints in this system.
The best-fit model indicates a visible/near-infrared flux ratio of $\sim2-5$
between the primary and secondary stars. If this is a long period binary, high resolution imaging observations may be
able to resolve it \citep[e.g.,][]{harris13}.

\citet{andrews19} demonstrate that Gaia astrometry can find hidden white dwarf companions at distances as far as several hundred parsecs.
In addition, Gaia can characterize orbits with periods ranging from 10 days to thousands of days. Using 0.1 mas as the size of the primary
star's orbit resolvable by Gaia \citep{andrews19} and a distance of 37 pc, Gaia should be able to easily resolve the
astrometric orbit for WD 1418$-$088 within the next several years.

\subsection{WD 1447$-$190}

\begin{figure}
\includegraphics[width=3.4in, bb=54 54 758 558]{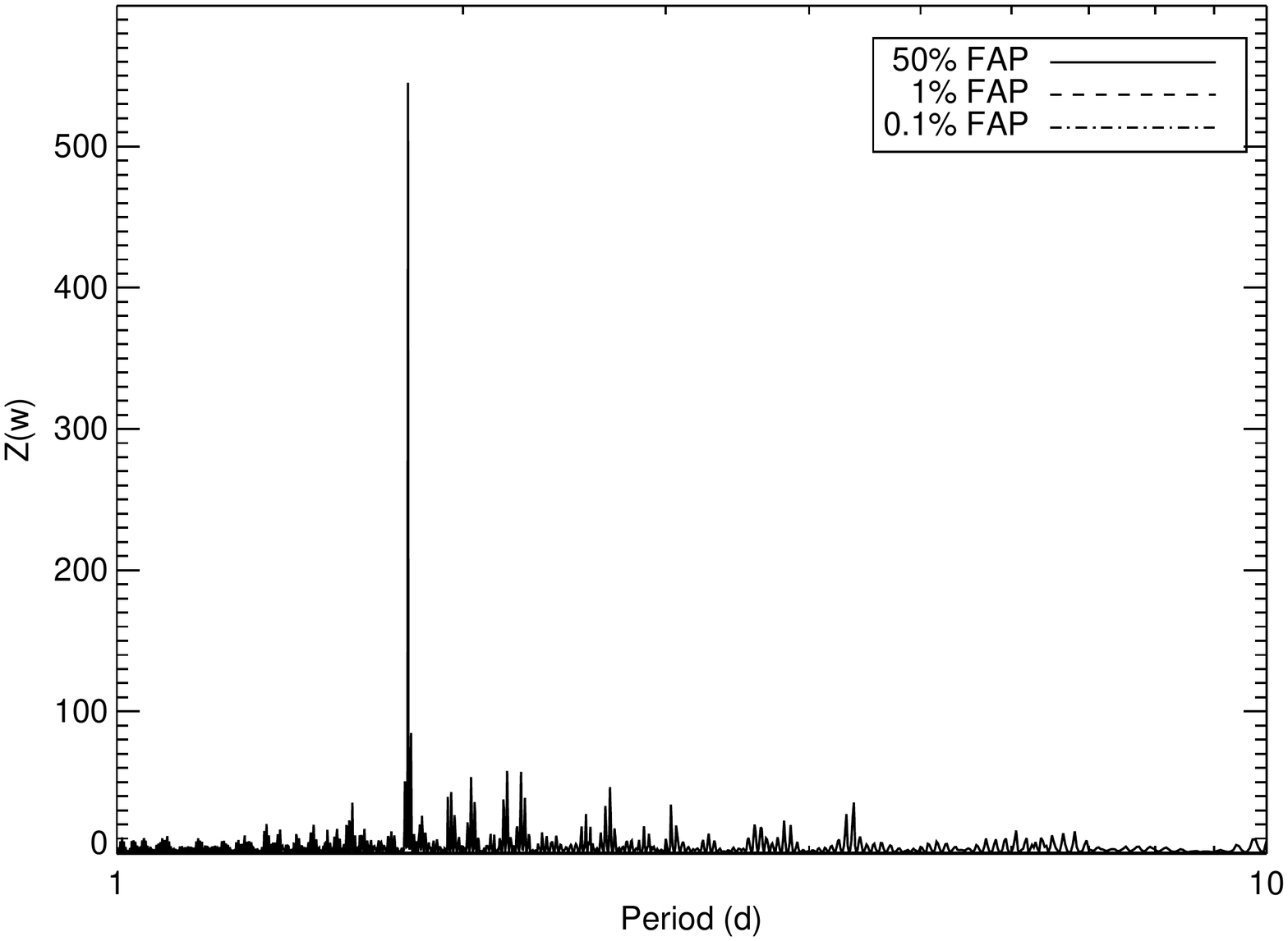}
\includegraphics[width=3.4in, bb=54 54 758 558]{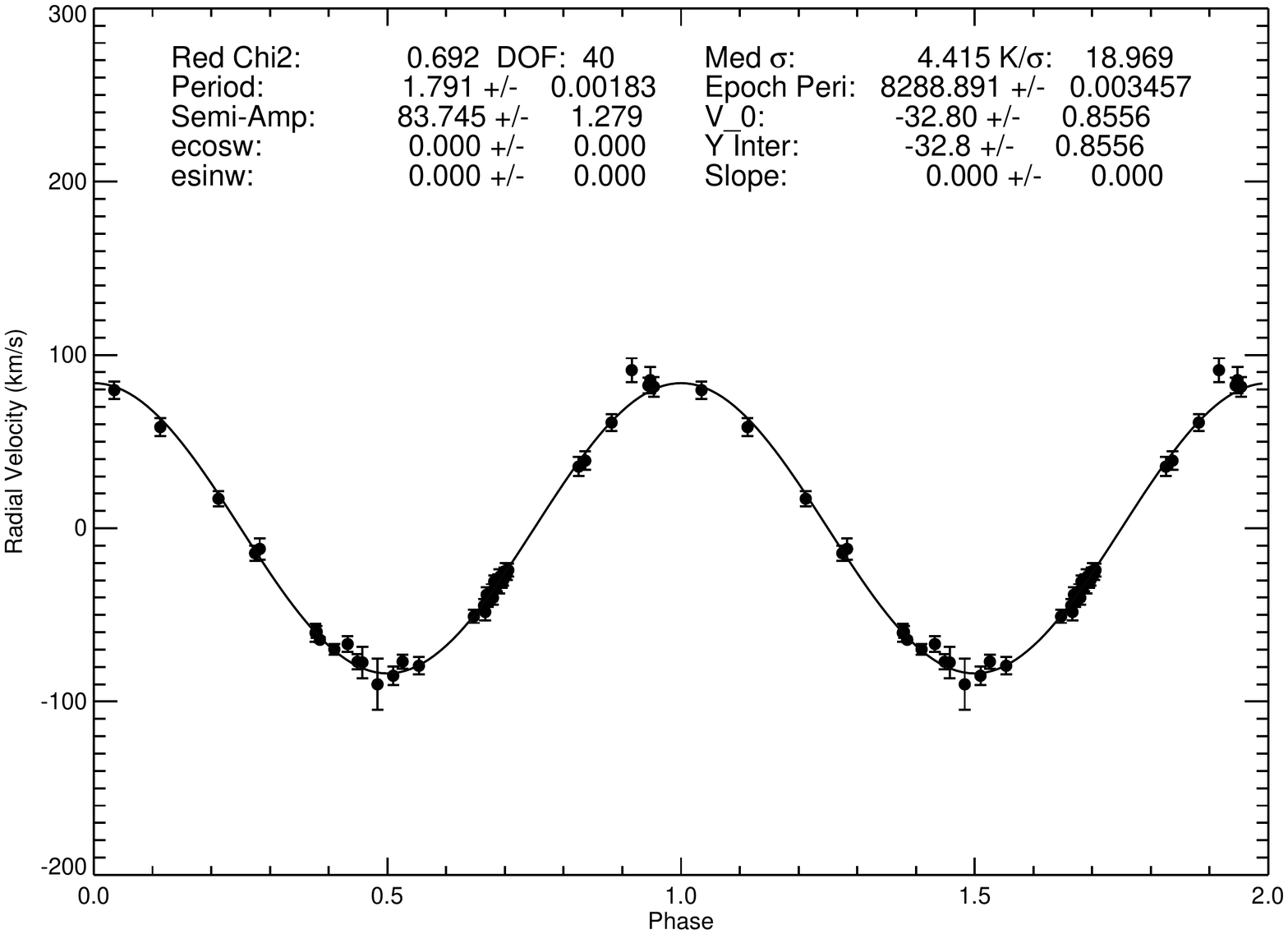}
\vspace{-0.1in}
\caption{{\it Top:} Lomb-Scargle periodogram for WD 1447$-$190. The period is well constrained and there are no significant
period aliases.
{\it Bottom:} Radial velocity measurements and the best-fitting orbital solution (solid line) for WD 1447$-$190
assuming a circular orbit. The best-fitting orbital period is 1.79 d.}
\label{fig:1447}
\end{figure}

Figure \ref{fig:1447} shows the radial velocity measurements and the Lomb-Scargle periodogram for
the single-lined binary WD 1447$-$190. There are no significant period aliases in the Lomb-Scargle diagram. We perform 1000 Monte Carlo
simulations for our orbital fits, and the results are consistent with the formal estimates from MPRVFIT. The best-fit orbital
solution has $P = 1.79083  \pm 0.00004$ d, $K = 83.8^{+1.2}_{-1.3}$ \kms, $\gamma = -32.7^{+0.8}_{-0.9}$ \kms, and a mass function of
$f= 0.109 \pm 0.005 \ M_{\odot}$.

Our best model-atmosphere fit to the Balmer lines and energy distribution of WD 1447$-$190 is presented in
Figure \ref{fig:fit3}. Given the lack of a complete orbital solution, all four atmospheric parameters are considered as
free parameters in the minimization process. We find $T_{\rm eff,1} = 8000 \pm 1000$ K, $\logg_1 = 7.65 \pm 0.20$ and
$T_{\rm eff,2} = 5000 \pm 500$ K, $\logg_2 = 7.50 \pm 0.20$, which convert to $M_1 = 0.41^{+0.10}_{-0.08} \ M_{\odot}$ and
$M_2 = 0.33^{+0.09}_{-0.07} \ M_{\odot}$ using our evolutionary sequences. Again, our composite model reproduces the spectroscopy
and photometry relatively well. For $M_1 = 0.41 M_{\odot}$, the mass function requires $M_2 \geq 0.42 M_{\odot}$, which suggests that
the secondary mass is closer to the upper limit of the mass range indicated by our model atmosphere analysis. 
Interestingly, our solution provides an elegant explanation for the single-lined nature of
this binary system: the secondary white dwarf has a very low effective temperature, hence its H$\alpha$ feature is simply too
weak to be observed.

\subsection{WD 1606+422}

\begin{figure}
\includegraphics[width=3.4in, bb=54 54 758 558]{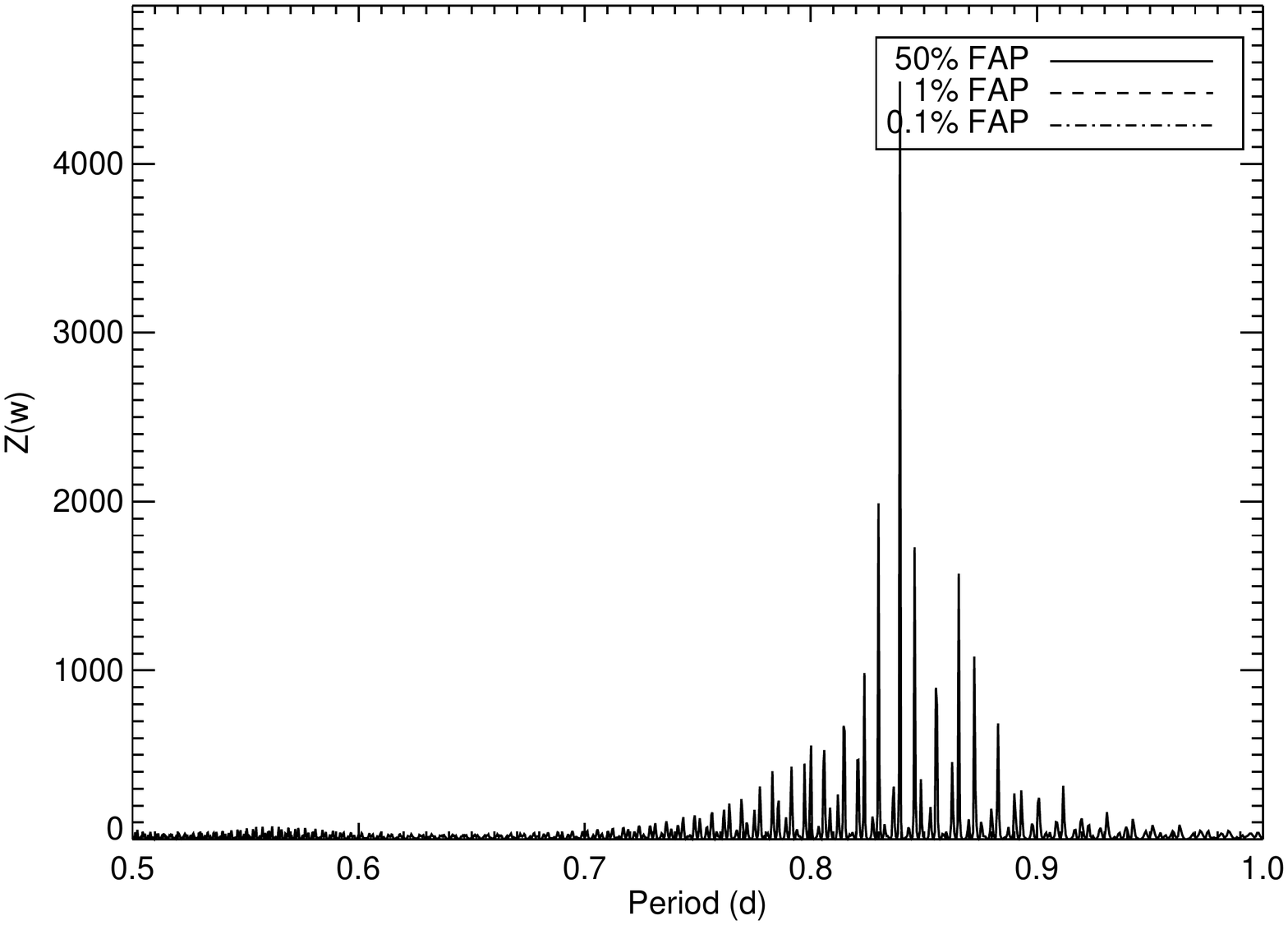}
\includegraphics[width=3.4in, bb=18 144 592 718]{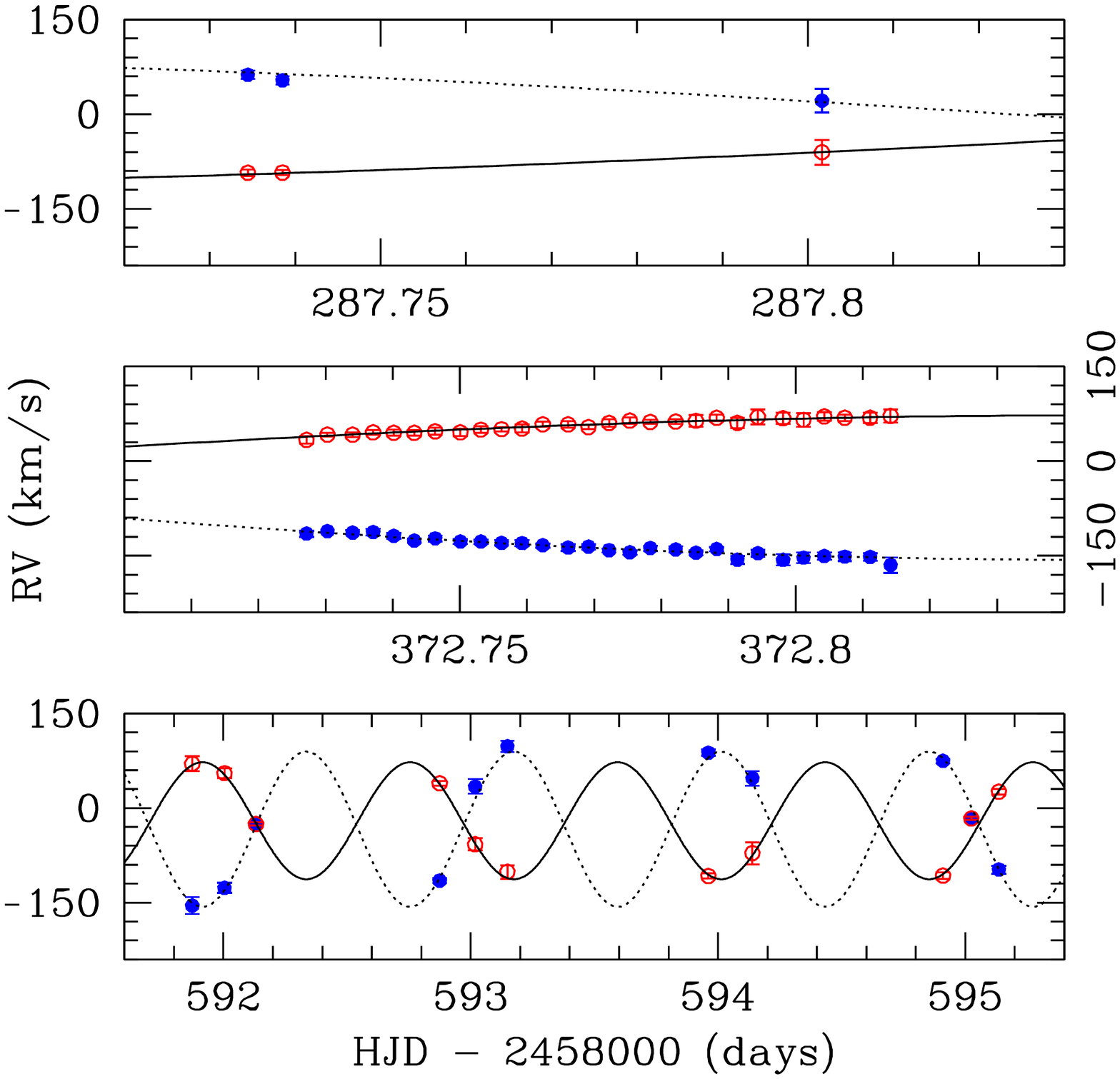}
\vspace{-0.1in}
\caption{{\it Top:} Lomb-Scargle periodogram for WD 1606+422. 
{\it Bottom:} Radial velocity measurements (open and filled points) and
the best-fitting orbital solutions (dotted and solid lines) for the two stars in WD 1606+422
assuming a circular orbit.}
\label{fig:1606}
\end{figure}

\begin{table*}
\caption{Orbital and physical parameters of the confirmed binary systems in the \citet{bedard17} sample of over-luminous white dwarfs.}
\begin{tabular}{cccccc}
\hline
Object & $\varpi$(mas) & P(days) & Masses ($M_{\odot}$) & Type & Reference\\
\hline
WD0135$-$052 &  79.21 $\pm$ 0.04 & 1.56 & 0.47 + 0.52 & SB2 & \citet{saffer88} \\
WD0311$-$649 &  27.29 $\pm$ 0.03  & 0.74 & 0.39 + 0.55 & SB2 & This paper \\
WD1242$-$105 &  24.80 $\pm$ 0.04 & 0.12 & 0.39 + 0.56 & SB2 & \citet{debes15}, \citet{subasavage17} \\ 
WD1606+422    &  23.07 $\pm$ 0.03 & 0.84 & 0.45 + 0.59 & SB2 & This paper \\
\hline
WD0101+048    &  44.86 $\pm$ 0.12 & $\sim6.4$ or 1.2 &  0.49 + ?             & SB1 & \citet{maxted00}\\
WD0326$-$273 & 43.43 $\pm$ 0.04 & 1.88 &  0.51 + $\geq0.59$  & SB1 & \citet{nelemans05}\\
WD1447$-$190 & 20.52 $\pm$ 0.05 & 1.79 &  0.41 + 0.33             & SB1 & This paper \\
WD1824+040    & 22.42 $\pm$ 0.09 & 6.27 &  0.43 + $\geq0.52$  & SB1 & \citet{morales05}\\
\hline
WD1639+153  & 31.48 $\pm$ 0.11 & 4 years &  0.93  + 0.91 (DA+DA?) & Astrometric & \citet{harris13}\\
                        &    &             &  0.98  + 0.69 (DA+DC?) &                    & \citet{harris13}\\
\hline
\end{tabular}
\end{table*}

Figure \ref{fig:1606} shows the radial velocity measurements and the Lomb-Scargle periodogram for
the double-lined spectroscopic binary WD 1606+422. Excluding the two spectra
where the H$\alpha$ lines from both stars overlap and appear as a single line, we have 40 velocity
measurements for each star. The best-fitting orbital solution has
$P = 0.83935 \pm 0.00002$  d, $K_1= 123.0 \pm 1.7$ \kms, $K_2 = 92.7 \pm 1.5$ \kms, 
$\gamma_1 = -33.6  \pm 1.5$ \kms, and a velocity offset of $\gamma_2 - \gamma_1 = 13.4 \pm 2.5$ \kms. 

The Lomb-Scargle periodogram shows that the period is well constrained for WD 1606+422. Performing
1000 Monte Carlo simulations, we derive
$P= 0.83935 \pm 0.00002$ d, $K_1 = 123.0^{+1.7}_{-1.8}$ \kms,
$K_2 = 92.8 \pm 1.7$ \kms, $\gamma_1 = -33.4^{+1.8}_{-1.9}$ \kms, 
$\gamma_2-\gamma_1 = 13.0^{+3.0}_{-3.2}$ \kms, and $\frac{K1}{K2} =1.33 \pm 0.03$.

Following our usual procedure, we find that there exists no solution satisfying all available constraints
(orbital parameters, spectroscopy, and photometry) simultaneously. However, a consistent solution is achievable
if the velocity offset is slightly smaller. Therefore, in what follows, we adopt
$\gamma_2-\gamma_1 = 11.0$ \kms, but keep the original confidence interval for the error propagation. Solving Equation
\ref{eq:mass} then yields $M_1 = 0.445^{+0.103}_{-0.039} \ M_{\odot}$ and $M_2 = 0.592^{+0.124}_{-0.045} \ M_{\odot}$.

Figure \ref{fig:fit3} displays our best model-atmosphere fit to the spectroscopic and photometric data of
WD 1606+422. As in the case of the other double-lined system WD 0311$-$649, we assume the surface gravities corresponding to
the masses derived from the orbital parameters, $\logg_1 = 7.70^{+0.19}_{-0.09}$ and $\logg_2 = 7.97^{+0.19}_{-0.08}$, while the effective temperatures are
determined from the fit. We obtain $T_{\rm eff,1} = 11,500 \pm 500$ K and $T_{\rm eff,2} = 13,300 \pm 500$ K, for which the Balmer
lines and the spectral energy distribution are reproduced well. 

\begin{figure}
\includegraphics[width=3.4in,clip=true,trim=1.7in 4.2in 1.7in 3.8in]{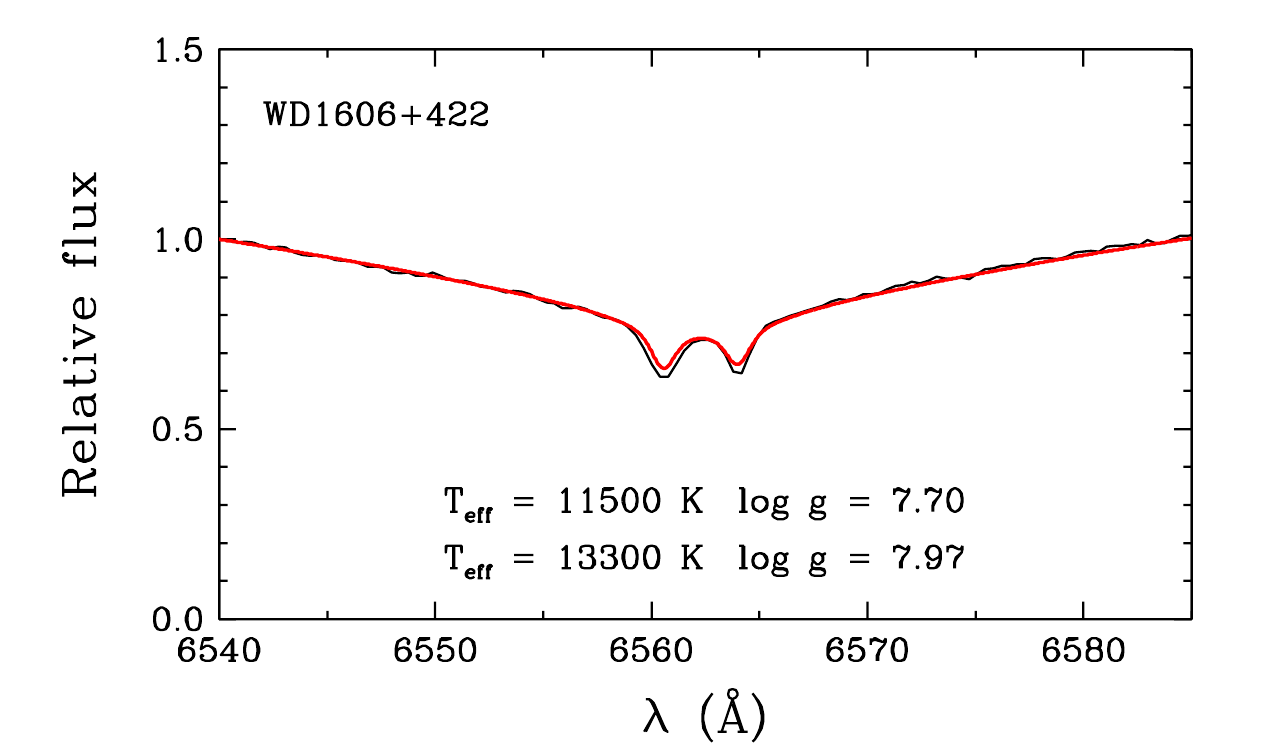}
\caption{Comparison of the observed double H$\alpha$ feature of WD 1606+422, shown as the black line, with that predicted by 
our best model-atmosphere fit, displayed as the red line.}
\label{fig:Ha1606}
\end{figure}

Figure \ref{fig:Ha1606} compares the observed double H$\alpha$ feature
of WD 1606+422 with that predicted by our best-fitting solution. As before, we improve the signal-to-noise by co-adding five of
our Gemini spectra, and we shift the individual synthetic spectra so that the positions of the observed and theoretical line
cores coincide. The agreement is quite good, although our model spectra appear slightly too shallow in the very core of the
lines. Finally, similar to WD 0311$-$649, our analysis suggests that the cooler white dwarf in WD 1606+422 falls within the
ZZ Ceti instability strip \citep{gianninas11}. Interestingly enough, \citet{gianninas11} and \citet{bognar18} reported
WD 1606+422 to be photometrically constant. However, it is possible that the luminosity variations have not been detected yet
because of their dilution by the light of the hotter component. Thus, we recommend that WD 1606+422 be further monitored for
photometric variability.

\section{Discussion}

\citet{bedard17} identified 15 over-luminous white dwarfs in their parallax sample that are inconsistent with being
single white dwarfs. Gaia DR2 parallaxes confirm the over-luminous nature of all but one of these targets, WD 1130+189.
Out of the 14 remaining binary candidates, one (WD 2048+809) lacks follow-up spectroscopy, four are SB2,
four are SB1, and one is a long-period astrometric binary. 

Table 1 presents the orbital parameters of all 9 confirmed binary systems in this sample.
The four SB2 systems have periods ranging from 0.12 to 1.56 d with mass ratios of 1.1-1.4, whereas the four SB1 systems have
periods in the range $\sim1-6$ d. The newly identified double-lined system WD 0311$-$649 is almost a twin of the WD 1242$-$105
binary, though with a much longer orbital period (0.74 d versus 0.12 d). WD 1639+153 is a $P=4$ year astrometric binary detected
in ground-based parallax observations by \citet{harris13}. Such long-period binary systems are extremely difficult to confirm by
radial velocity observations, and are likely hiding in over-luminous white dwarf samples.

Four of the targets in our sample have follow-up radial velocity observations which effectively rule out short-period binary systems.
In addition to WD 1418$-$088 discussed above, WD0126+101, WD0142+312,  and WD2111+261 \citep{maxted00,napiwotzki19}
have 8-12 radial velocity observations that do not reveal any significant variability. However, all of these targets appear
over-luminous based on their parallax measurements. In addition, atmospheric model fits to their spectra suggest average mass
white dwarfs, whereas fits to their photometry indicate much lower masses. 

For example, WD 0142+312 has the best-fit spectroscopic
estimates of $T_{\rm eff} = 9270 \pm 130$ K and $\log{g}=8.12 \pm 0.05$ \citep{limoges15}, whereas the photometric fit indicates
$T_{\rm eff} = 8790$ K and $\log{g}=7.54$. Both WD 0142+312 and WD 2111+261 are inconsistent with being single stars
at the $>5\sigma$ level. The case for a binary system is weaker for WD 0126+101, as the best-fit estimates from spectroscopy
and photometry differ by only $2\sigma$. Nevertheless, it is clear that at least three (WD 0142+312, WD 1418$-$088, and WD 2111+261),
and perhaps all four of these objects without any significant radial velocity variations must be binary white dwarfs. 

Our observations, as well as the previous radial velocity measurements in the literature, of these targets are not sensitive to
month and year-long orbital periods, and would not be able to detect long period systems like the astrometric binary
WD 1639+153 \citep{harris13}. Hence, these four targets are likely long period binary systems that can be confirmed through
either high-resolution imaging observations or Gaia astrometry \citep{andrews19}.

It is interesting to compare our binary sample to that of the larger sample from the SPY survey. \citet{napiwotzki19} identified
39 double degenerate binaries, half of which are SB2 systems. 
However, they found significant radial velocity variations in only 16 of the 44 low-mass ($M\leq0.45 M_{\odot}$) white
dwarfs in their sample, indicating that either these low-mass white dwarfs are single or that they have substellar mass
companions. Given that the mean detection efficiency of the SPY survey degrades quickly for month and longer timescales
\citep[][see their Fig. 6]{napiwotzki19}, it is likely that a significant fraction of these are long period binary white dwarfs with
higher masses. For example, two of the over-luminous white dwarfs in the \citet{bedard17} sample, WD 0126+101 and WD 1418$-$088,
are included in the SPY sample as single low-mass white dwarfs, but our analysis suggests that they are instead long period binary
systems of more massive white dwarfs.

\begin{figure}
\centering
\includegraphics[width=3.2in, bb=18 144 592 718]{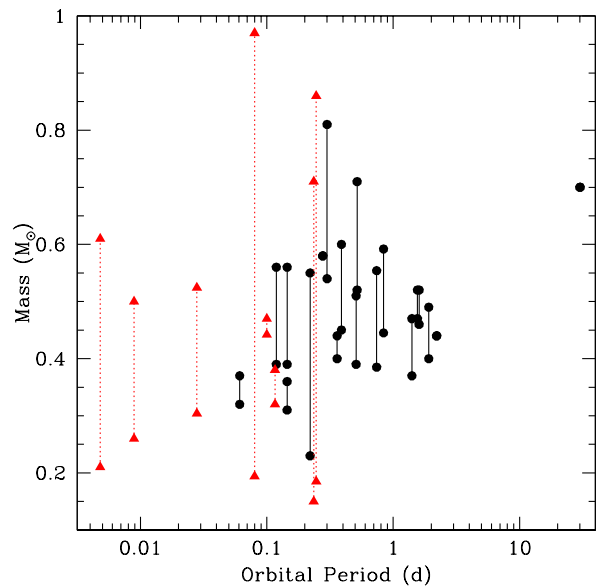}
\caption{Mass and orbital period distribution of all known SB2 (circles) and eclipsing (triangles) double white dwarfs with orbital constraints.
The lines connect the components of each binary.}
\label{fig:dist}
\end{figure}

Figure \ref{fig:dist} shows the mass and orbital period distribution of all known double-lined spectroscopic binary white dwarfs
and eclipsing double white dwarfs. The lines connect the components of each binary. We limit this
figure to only SB2 and eclipsing systems, where the component masses can be
constrained reliably. There are 19 SB2 white dwarfs, including the two newly identified
systems presented in this work, and 8 eclipsing systems \citep[][and references therein]{hallakoun16,burdge19}. 
The latter are found at short orbital periods ($P\leq0.25$ d), and dominated by ELM white dwarfs,
which usually have relatively massive companions \citep{andrews14,boffin15,brown17}.
On the other hand, the double-lined binaries are nearly equal-mass ratio systems. This is not surprising as the spectral lines from
both stars would be visible only if both have comparable luminosities, which depend on the radii, and therefore the masses
of the two white dwarfs in the system. With further discoveries of double-lined systems among the over-luminous white dwarf population
in Gaia \citep{marsh19}, we may finally be able to have a large enough sample of SB2 systems to compare against and constrain
the population synthesis models for double white dwarfs \citep[e.g.,][]{nelemans01}.

\section{Conclusions}

We presented follow-up spectroscopy of four over-luminous white dwarfs identified by \citet{bedard17}. Two of these, WD 0311$-$649
and WD 1606+422, are double-lined systems, and we provide orbital periods, mass ratios, component masses, and effective
temperatures of each binary based on their spectral line profiles, spectral energy distributions, and radial velocity data.  
An additional system, WD 1447$-$190, is a single-lined binary with a period of 1.79 d, whereas WD 1418$-$088 does not
show any significant velocity variations. 

Studying the 15 over-luminous white dwarfs in the \citet{bedard17} sample, and ignoring WD 2048+809 (no follow-up spectroscopy)
and WD 1130+189 (not over-luminous based on Gaia DR2 parallax), we find that four are SB2, four are SB1, one is an
astrometric binary, and four appear to show no significant radial velocity variations.
However, there are significant discrepancies between the spectroscopic and photometric fits for the latter four stars, and the only
way to resolve this issue is if they are in long period binary systems. Follow-up high spatial resolution imaging and/or Gaia
astrometry \citep{andrews19} may resolve these four systems. We also argue that the over-abundance of single low-mass
white dwarfs in the SPY survey \citep{napiwotzki19} is likely due to a similar problem, and that at least some of those objects
are likely in long period binary systems as well.

Our results provide strong evidence that all 13 of these over-luminous white dwarfs with follow-up spectroscopy are indeed
double degenerates. In addition, one of these systems, WD 1242$-$105 has a short enough orbital period to merge within a Hubble time
\citep{debes15}. However, we refrain from discussing the implications of our results on the overall space density and merger rate of
double white dwarfs due to biases in our sample selection. The 219 stars analyzed in \citet{bedard17} were selected simply because
they had a parallax measurement available at the time. They also used parallax measurements from several different sources
(e.g., Hipparcos, Yale Parallax Catalog, USNO, Gaia DR1). Hence, our over-luminous white dwarf sample has strong selection
effects and is not statistically complete.

\section*{Acknowledgements}

We thank the referee, Dr. Jeff Andrews, for a constructive report.
We gratefully acknowledge the support of the NSF under grant AST-1906379.
This work is supported in part by the NSERC Canada and by the Fund FRQ-NT (Qu\'ebec).
The authors wish to recognize and acknowledge the very significant cultural role and reverence
that the summit of Mauna Kea has always had within the indigenous Hawaiian community. We are
most fortunate to have the opportunity to conduct observations from this mountain.

This work benefited from a workshop held at DARK in July 2019 that was funded
by the Danish National Research Foundation (DNRF132). We thank Dr. Josiah
Schwab for his efforts in organising this workshop.

Based on observations obtained at the Gemini Observatory, which is operated by the Association of Universities for Research in Astronomy, Inc., under a cooperative agreement with the NSF on behalf of the Gemini partnership: the National Science Foundation (United States), National Research Council (Canada), CONICYT (Chile), Ministerio de Ciencia, Tecnolog\'{i}a e Innovaci\'{o}n Productiva (Argentina), Minist\'{e}rio da Ci\^{e}ncia, Tecnologia e Inova\c{c}\~{a}o (Brazil), and Korea Astronomy and Space Science Institute (Republic of Korea).

This work was supported by a NASA Keck PI Data Award, administered by the NASA Exoplanet
Science Institute. Data presented herein were obtained at the W. M. Keck Observatory from
telescope time allocated to the National Aeronautics and Space Administration through the
agency's scientific partnership with the California Institute of Technology and the University of
California. The Observatory was made possible by the generous financial support of the W. M.
Keck Foundation.

Based on observations obtained at the Southern Astrophysical Research (SOAR) telescope, which is a joint project of the Minist\'{e}rio da Ci\^{e}ncia, Tecnologia, Inova\c{c}\~{o}es e Comunica\c{c}\~{o}es (MCTIC) do Brasil, the U.S. National Optical Astronomy Observatory (NOAO), the University of North Carolina at Chapel Hill (UNC), and Michigan State University (MSU).

\bibliography{master}
\bsp
\label{lastpage}

\appendix
\section{Radial Velocity Data}

\begin{table}
\scriptsize
\centering
\caption{WD 0311-649}
\begin{tabular}{crr}
\hline
HJD$-$2450000 & $V1_{helio}$ & $V2_{helio}$ \\
(days) & (\kms) & (\kms)\\
\hline
8393.74603027 & $0.0 \pm 3.9 $ & $133.7 \pm 4.0$ \\
8393.74800758 & $-5.3 \pm 7.3 $ & $134.8 \pm 4.8$ \\
8393.75141472 & $-4.2 \pm 6.2 $ & $137.1 \pm 5.4$ \\
8393.75339236 & $3.5 \pm 10.2 $ & $133.4 \pm 5.7$ \\
8393.75536627 & $0.2 \pm 7.7 $ & $141.4 \pm 6.4$ \\
8393.75734092 & $1.3 \pm 7.2 $ & $138.3 \pm 5.5$ \\
8393.75931542 & $5.6 \pm 11.3 $ & $135.8 \pm 9.5$ \\
8393.76129677 & $8.9 \pm 8.2 $ & $138.1 \pm 5.3$ \\
8393.76326492 & $3.3 \pm 7.4 $ & $131.4 \pm 5.3$ \\
8393.76619544 & $-2.6 \pm 7.9 $ & $133.0 \pm 5.3$ \\
8393.76817239 & $7.2 \pm 5.7 $ & $135.1 \pm 4.3$ \\
8393.77014865 & $5.2 \pm 8.1 $ & $134.0 \pm 6.2$ \\
8393.77213876 & $1.6 \pm 8.7 $ & $140.4 \pm 7.5$ \\
8393.77411675 & $11.0 \pm 10.5 $ & $140.1 \pm 7.2$ \\
8393.77608770 & $-0.7 \pm 8.9 $ & $125.3 \pm 7.9$ \\
8393.77806197 & $-3.5 \pm 4.9 $ & $128.8 \pm 5.0$ \\
8393.78099341 & $-4.8 \pm 6.6 $ & $129.4 \pm 4.9$ \\
8393.78297661 & $3.4 \pm 7.0 $ & $136.4 \pm 4.9$ \\
8393.78494521 & $5.5 \pm 7.7 $ & $128.0 \pm 5.5$ \\
8393.78691986 & $10.0 \pm 7.1 $ & $133.4 \pm 4.8$ \\
8393.78889460 & $4.2 \pm 10.9 $ & $127.6 \pm 5.1$ \\
8393.79087339 & $3.6 \pm 8.2 $ & $130.9 \pm 6.1$ \\
8393.79284709 & $10.1 \pm 5.4 $ & $128.8 \pm 3.9$ \\
8393.79576315 & $0.4 \pm 6.8 $ & $130.4 \pm 5.8$ \\
8393.79773732 & $14.9 \pm 12.8 $ & $131.9 \pm 8.3$ \\
8393.79971394 & $13.0 \pm 6.1 $ & $132.4 \pm 4.4$ \\
8393.80169190 & $3.7 \pm 8.8 $ & $127.9 \pm 4.9$ \\
8393.80366792 & $4.5 \pm 7.7 $ & $123.9 \pm 5.2$ \\
8393.80564173 & $7.6 \pm 5.6 $ & $120.7 \pm 4.8$ \\
8393.80761778 & $18.7 \pm 10.3 $ & $132.2 \pm 5.3$ \\
8393.81101586 & $12.4 \pm 6.4 $ & $122.1 \pm 5.3$ \\
8393.81299096 & $8.3 \pm 11.5 $ & $122.5 \pm 8.8$ \\
8393.81496490 & $14.1 \pm 8.5 $ & $122.5 \pm 9.6$ \\
8393.81694324 & $-0.9 \pm 9.2 $ & $117.8 \pm 7.3$ \\
8393.81891741 & $0.4 \pm 8.9 $ & $117.4 \pm 6.2$ \\
8393.82089272 & $4.8 \pm 10.8 $ & $115.2 \pm 7.2$ \\
8393.82286781 & $13.7 \pm 10.0 $ & $117.3 \pm 8.3$ \\
8393.82578447 & $3.3 \pm 9.8 $ & $118.3 \pm 6.0$ \\
8393.82775933 & $5.8 \pm 6.3 $ & $119.1 \pm 7.5$ \\
8393.82973347 & $9.7 \pm 8.5 $ & $116.8 \pm 7.3$ \\
8393.83172374 & $12.0 \pm 11.7 $ & $111.9 \pm 7.1$ \\
8393.83369756 & $10.7 \pm 11.1 $ & $104.8 \pm 5.7$ \\
8393.83568803 & $16.5 \pm 8.7 $ & $108.8 \pm 7.1$ \\
8393.83767901 & $-1.0 \pm 21.2 $ & $97.1 \pm 9.7$ \\
8393.84061334 & $11.5 \pm 17.6 $ & $106.2 \pm 12.2$ \\
8393.84258957 & $15.3 \pm 7.9 $ & $103.6 \pm 7.0$ \\
8393.84456527 & $23.9 \pm 11.4 $ & $106.5 \pm 11.1$ \\
8393.84655678 & $16.4 \pm 20.3 $ & $101.5 \pm 12.2$ \\
8393.84853167 & $11.4 \pm 12.6 $ & $97.4 \pm 8.5$ \\
8393.85050650 & $22.0 \pm 8.5 $ & $99.2 \pm 11.6$ \\
8393.85248162 & $21.3 \pm 12.4 $ & $105.2 \pm 13.2$ \\
8393.85540216 & $5.2 \pm 25.6 $ & $92.3 \pm 15.2$ \\
8393.85737970 & $16.2 \pm 29.0 $ & $88.6 \pm 19.3$ \\
8393.85936985 & $25.3 \pm 12.3 $ & $99.8 \pm 11.3$ \\
8393.86134366 & $11.2 \pm 12.6 $ & $84.8 \pm 9.8$ \\
8393.86333390 & $24.7 \pm 13.2 $ & $93.8 \pm 8.9$ \\
8393.86532509 & $25.3 \pm 13.5 $ & $95.5 \pm 12.4$ \\
8796.50239379 & $114.0 \pm 7.1 $ & $-24.3 \pm 7.5$ \\
8796.64593769 & $56.3 \pm 5.5 $ & $56.3 \pm 5.5$ \\
8796.79867849 & $3.3 \pm 9.7 $ & $127.6 \pm 7.9$ \\
8797.57631070 & $14.1 \pm 9.5 $ & $131.7 \pm 8.0$ \\
8797.67691008 & $56.1 \pm 15.0 $ & $97.2 \pm 10.5$ \\
8797.86212509 & $116.4 \pm 6.6 $ & $-25.8 \pm 4.3$ \\
8798.51407694 & $57.5 \pm 11.4 $ & $11.7 \pm 7.6$ \\
8798.70385557 & $117.1 \pm 8.2 $ & $-28.2 \pm 6.4$ \\
8798.81964492 & $46.5 \pm 3.4 $ & $46.5 \pm 3.4$ \\
8799.51585147 & $96.1 \pm 12.8 $ & $-3.3 \pm 13.1$ \\
8799.75808955 & $1.7 \pm 7.7 $ & $129.0 \pm 5.8$ \\
8799.76176339 & $-1.8 \pm 10.0 $ & $135.1 \pm 8.5$ \\
8801.51700741 & $103.5 \pm 12.3 $ & $-10.6 \pm 9.3$ \\
8801.79271799 & $ 51.0 \pm 2.2 $ & $51.0 \pm 2.2$ \\
\hline
\end{tabular}
\end{table}

\begin{table}
\centering
\caption{WD 1418$-$088}
\begin{tabular}{cr}
\hline
HJD$-$2450000 & $V_{helio}$ \\
(days) & (\kms) \\
\hline
1739.5605 & $-37.1 \pm 1.5$ \\
1742.5722 & $-30.9 \pm 1.6$ \\
8287.75722641 & $-31.3 \pm 2.3$ \\
8287.82388136 & $-28.2 \pm 2.4$ \\
8300.79887603 & $-31.1 \pm 3.9$ \\
8300.80257108 & $-35.1 \pm 4.1$ \\
8300.80700933 & $-28.2 \pm 3.7$ \\
8300.81070462 & $-32.3 \pm 3.1$ \\
8300.81440187 & $-28.8 \pm 4.0$ \\
8300.81809752 & $-34.2 \pm 2.6$ \\
8300.82252975 & $-27.2 \pm 4.8$ \\
8300.82622664 & $-34.2 \pm 3.1$ \\
8300.82992265 & $-27.7 \pm 4.8$ \\
8300.83361871 & $-39.1 \pm 6.0$ \\
8300.83829250 & $-31.1 \pm 4.4$ \\
8300.84198874 & $-37.6 \pm 3.4$ \\
8300.84568688 & $-33.6 \pm 4.2$ \\
8300.84938494 & $-32.6 \pm 4.6$ \\
8300.85382737 & $-36.8 \pm 3.4$ \\
8300.85752113 & $-36.8 \pm 3.0$ \\
8300.86121842 & $-35.1 \pm 3.5$ \\
8300.86491463 & $-28.6 \pm 3.9$ \\
8300.86934108 & $-34.6 \pm 4.3$ \\
8300.87303809 & $-33.2 \pm 4.3$ \\
\hline
\end{tabular}
\end{table}

\begin{table}
\centering
\caption{WD 1447$-$190}
\begin{tabular}{cr}
\hline
HJD$-$2450000 & $V_{helio}$ \\
(days) & (\kms) \\
\hline
8287.79004386 & $-97.1 \pm 1.5$ \\
8287.83398725 & $-102.6 \pm 3.1$ \\
8309.78038901 & $-77.4 \pm 3.8$ \\
8309.78408475 & $-81.1 \pm 4.8$ \\
8309.78851329 & $-71.1 \pm 4.4$ \\
8309.79220918 & $-74.4 \pm 3.9$ \\
8309.79590724 & $-71.6 \pm 4.6$ \\
8309.79960426 & $-69.4 \pm 2.6$ \\
8309.80402782 & $-70.0 \pm 4.9$ \\
8309.80771825 & $-72.9 \pm 4.1$ \\
8309.81141411 & $-63.6 \pm 3.4$ \\
8309.81511482 & $-66.0 \pm 4.0$ \\
8309.81953192 & $-65.9 \pm 3.2$ \\
8309.82322608 & $-63.0 \pm 3.8$ \\
8309.82692321 & $-63.5 \pm 7.0$ \\
8309.83062011 & $-60.4 \pm 4.0$ \\
8309.83504889 & $-63.4 \pm 3.7$ \\
8309.83874442 & $-62.9 \pm 2.9$ \\
8309.84244060 & $-58.3 \pm 2.8$ \\
8309.84613589 & $-57.4 \pm 4.5$ \\
8309.85055716 & $-58.7 \pm 3.8$ \\
8309.85425293 & $-57.0 \pm 3.8$ \\
8544.82924001 & $58.4 \pm 6.8$ \\
8544.88585415 & $52.7 \pm 7.5$ \\
8545.75376632 & $-99.6 \pm 4.6$ \\
8545.79880531 & $-110.2 \pm 9.1$ \\
8545.84483738 & $-122.8 \pm 14.8$ \\
8545.89278086 & $-117.9 \pm 5.3$ \\
8546.83294299 & $46.9 \pm 5.0$ \\
8588.63922836 & $-92.5 \pm 3.4$ \\
8588.76353703 & $-109.8 \pm 4.4$ \\
8588.90132700 & $-109.7 \pm 4.1$ \\
8602.77792314 & $-47.2 \pm 4.4$ \\
8602.79230148 & $-44.7 \pm 6.1$ \\
8603.86506466 & $28.2 \pm 4.8$ \\
8605.55550400 & $2.8 \pm 5.5$ \\
8605.76824108 & $49.6 \pm 4.5$ \\
8606.54405099 & $-93.1 \pm 5.1$ \\
8606.85924347 & $-112.1 \pm 5.0$ \\
8616.52990044 & $48.9 \pm 5.6$ \\
8616.81562358 & $25.6 \pm 5.1$ \\
8636.69212753 & $-15.7 \pm 4.3$ \\
8637.47080183 & $-83.7 \pm 3.7$ \\
8637.81049184 & $6.2 \pm 5.4$ \\
\hline
\end{tabular}
\end{table}

\begin{table}
\centering
\caption{WD 1606+422}
\begin{tabular}{crr}
\hline
HJD$-$2450000 & $V1_{helio}$ & $V2_{helio}$ \\
(days) & (\kms) & (\kms)\\
\hline
8287.73442998 & $62.1 \pm 6.7$ & $-92.9 \pm 5.4$ \\
8287.73849062 & $53.7 \pm 6.7$ & $-92.6 \pm 4.2$ \\
8287.80162519 & $21.6 \pm 18.4$ & $-60.4 \pm 19.5$ \\
8372.72714356 & $-115.1 \pm 4.9$ & $33.9 \pm 6.0$ \\
8372.73020305 & $-111.2 \pm 3.0$ & $41.7 \pm 2.9$ \\
8372.73398796 & $-113.2 \pm 3.1$ & $41.8 \pm 3.8$ \\
8372.73704803 & $-112.7 \pm 5.2$ & $45.7 \pm 3.8$ \\
8372.74011007 & $-118.4 \pm 4.4$ & $45.0 \pm 4.8$ \\
8372.74316907 & $-126.0 \pm 3.0$ & $45.1 \pm 4.4$ \\
8372.74623077 & $-122.7 \pm 5.1$ & $47.3 \pm 5.7$ \\
8372.75003536 & $-127.5 \pm 4.0$ & $45.6 \pm 5.9$ \\
8372.75309717 & $-127.8 \pm 3.2$ & $49.8 \pm 5.1$ \\
8372.75615667 & $-129.7 \pm 4.0$ & $50.4 \pm 3.3$ \\
8372.75921767 & $-130.2 \pm 4.7$ & $51.7 \pm 6.4$ \\
8372.76227795 & $-133.7 \pm 3.4$ & $58.4 \pm 4.1$ \\
8372.76606043 & $-137.4 \pm 5.0$ & $58.2 \pm 4.9$ \\
8372.76911994 & $-136.0 \pm 4.3$ & $54.0 \pm 3.3$ \\
8372.77217918 & $-141.4 \pm 3.1$ & $60.3 \pm 6.1$ \\
8372.77524041 & $-144.9 \pm 3.5$ & $64.0 \pm 5.0$ \\
8372.77830567 & $-138.1 \pm 4.2$ & $62.2 \pm 3.6$ \\
8372.78208815 & $-140.1 \pm 4.7$ & $62.7 \pm 3.9$ \\
8372.78514742 & $-145.3 \pm 3.9$ & $63.9 \pm 9.5$ \\
8372.78820890 & $-139.6 \pm 3.3$ & $68.4 \pm 5.2$ \\
8372.79126849 & $-156.7 \pm 6.3$ & $60.3 \pm 8.1$ \\
8372.79432937 & $-146.1 \pm 5.0$ & $70.0 \pm 11.8$ \\
8372.79810607 & $-156.5 \pm 8.6$ & $67.9 \pm 8.6$ \\
8372.80116626 & $-153.1 \pm 8.8$ & $65.3 \pm 10.6$ \\
8372.80422631 & $-150.8 \pm 5.8$ & $70.7 \pm 6.6$ \\
8372.80728686 & $-151.7 \pm 7.0$ & $68.7 \pm 5.3$ \\
8372.81106865 & $-152.2 \pm 6.1$ & $68.7 \pm 7.9$ \\
8372.81412870 & $-165.0 \pm 11.9$ & $71.5 \pm 10.6$ \\
8591.87407156 & $-154.6 \pm 13.5$ & $70.2 \pm 12.0$ \\
8592.00464574 & $-126.5 \pm 8.1$ & $55.2 \pm 7.8$ \\
8592.13206901 & $-25.2 \pm 2.8$ & $-25.2 \pm 2.8$ \\
8592.87443003 & $-115.1 \pm 3.9$ & $39.0 \pm 3.5$ \\
8593.01822250 & $34.1 \pm 11.5$ & $-57.6 \pm 9.6$ \\
8593.14855262 & $97.7 \pm 8.9$ & $-101.3 \pm 10.8$ \\
8593.96107311 & $87.5 \pm 4.8$ & $-107.7 \pm 4.2$ \\
8594.13898214 & $47.1 \pm 11.7$ & $-71.8 \pm 17.4$ \\
8594.90966600 & $74.8 \pm 3.3$ & $-107.2 \pm 5.3$ \\
8595.02447094 & $-16.3 \pm 2.4$ & $-16.3 \pm 2.4$ \\
8595.13547162 & $-97.8 \pm 6.4$ & $25.7 \pm 4.6$ \\
\hline
\end{tabular}
\end{table}

\end{document}